\documentclass[journal,transmag]{IEEEtran}
\ifCLASSINFOpdf
\else
\fi

\usepackage{ccaption}
\usepackage{amsmath}
\usepackage{amsbsy}
\usepackage{dcolumn}
\usepackage{multirow}
\usepackage[table]{xcolor}
\usepackage{wrapfig}
\usepackage{pifont}
\usepackage{float}
\usepackage{adjustbox}
\usepackage{wrapfig}
\usepackage{soul}
\usepackage{footnote}
\usepackage{times}
\usepackage{lipsum}

\usepackage[noadjust]{cite}

\usepackage[T1]{fontenc}
\usepackage{textcomp}
\usepackage{graphicx}
\usepackage{graphics}

\usepackage{caption}
\usepackage{subcaption}

\usepackage{amssymb,amstext,amsmath}
\usepackage{float}
\usepackage{booktabs}
\usepackage{array,ragged2e}
\usepackage[table]{xcolor}
\usepackage{multirow}
\usepackage{algorithm}
\usepackage{algpseudocode}
\usepackage{pifont}
\begin{document}
%
\title{Gaussian Processes for Traffic Speed Prediction at Different Aggregation Levels}
%
%
%

\author{Gurcan~Comert
\thanks{G. Comert is with the Department
of Computer Science, Physics, and Engineering, Benedict College, Columbia,
SC, 29204 USA e-mail: gurcan.comert@benedict.edu. He is also adjunct researcher at Information Trust Institute, University of Illinois, Urbana-Champaign, IL 61801 USA}
}

\maketitle

\begin{abstract}
Dynamic behavior of traffic adversely affect the performance of the prediction models in intelligent transportation applications. This study applies Gaussian processes (GPs) to traffic speed prediction. Such predictions can be used by various transportation applications, such as real-time route guidance, ramp metering, congestion pricing and special events traffic management. One-step predictions with various aggregation levels (1 to 60-minute) are tested for performance of the generated models. Univariate and multivariate GPs are compared with several other linear, nonlinear time series, and Grey system models using loop and Inrix probe vehicle datasets from California, Portland, and Virginia freeways respectively. Based on the test data samples, results are promising that GP models are able to consistently outperform compared models with similar computational times.
\end{abstract}
%
\IEEEpeerreviewmaketitle
\section{Introduction}
%
%
%
%
\IEEEPARstart{M}{ajority} of the critical infrastructures today are impacted by the improved surveillance technologies. Such systems are generating various data types: user presence via inductance loops, radar, video cameras, and connected vehicles. These data are then input for many intelligent transportation systems (ITS) such as advanced traveler information systems, real-time route guidance, and emergency response systems to improve safety and mobility of people and goods. The data in many cases partially covers the observed system, that is, a real-time traffic parameter prediction mechanism is required. Accurate and robust prediction of traffic parameters (e.g., average speed, travel time, volume, and occupancy) at various locations, facility types, and changing conditions is a critical problem that any improvement would yield more efficient operations management and control strategies. For instance, better real-time travel time prediction can result in better shipment scheduling or more accurately finding the fastest way to a destination for delivering goods \cite{Cheng20128356}. Even though system surveillance is becoming available, transportation and its connected systems are subjected to show unexpected or expected uncertainties due to various external factors such as incidents, inclement weather, special events, driver behaviors, and cyber attacks that adversely affect the performance and accuracy of prediction models. This paper develops and tests online adaptive methods based on Gaussian Processes (GPs) for reliable and robust short-to-midterm (i.e., 1 to 60 minutes (min)) traffic predictions. The macroscopic traffic parameters investigated in this paper are average speed, travel time, volume, and occupancy for different aggregation levels (1-60 min). These parameters are selected since (1) they are key parameters highly impacted by the aforementioned external factors as well as covariates such as weather, demand surges, driver characteristics, road work, and roadway geometry information that are not necessarily stamped together with observed data and (2) they are critical parameters used in wide range of traffic management and control, level of service analysis, planning, and safety applications at transportation networks. Gaussian Processes provide distributions over models that may better forecast parameters that are impacted by the nonlinearities. This approach has been utilized by many researchers for time series modeling \cite{clements2010forecast}. However, traffic data generation processes (DGPs) are not necessarily independent and identically distributed (i.i.d) due to different locations or sudden shifts that lead to non-stationarity and nonlinearity. Moreover, underlying distributions for DGPs are not always known. Hence, prediction models need to be responsive to possible changed dynamics via updating model parameters (i.e., retraining) ideally inherently with low amount of data and computational complexity.  

As one of the machine learning regression techniques, Gaussian Processes are fundamentally researched in may fields \cite{rasmussen2003gaussian} especially wireless network traffic \cite{bayati2020gaussian}. In traffic prediction problems, they are applied to flow prediction \cite{xie2010gaussian,sun2010variational}. These early studies are rightfully more towards understanding this relatively new framework. They were not extensively compared against similar competing methods. Xie et al. \cite{xie2010gaussian} compared GPs against linear time series models obtaining similar results to this study. As GPs are also referred as kriging, in this setting it was applied in \cite{le2016local}. Question of update and generalization in classical regression set up are still concept of investigation topics. Performance of GPs would show similar as higher order polynomial approximations. Moreover, for spatial data in kriging application missing data GPs were also applied for data imputation \cite{wang2018short}. In presence of explanatory variables, indeed GPs are able to accurately predict missing data. However, with only spatial correlations and categorical tags, performances are very similar to K-means or other clustering algorithms. Most similar paper to this study is from wireless network prediction \cite{bayati2020gaussian}. Authors compared GP-based regressions against time series and neural networks (NN) over different aggregation levels up to 30 minutes. Authors show better accuracy of time series and NNs.     

Grey System theory applications in transportation, the applications of the theory have been on volume, traffic accident, and pavement design data~\cite{an2012exploring,gao2010road,liu2014highway,jiang2005gray, du2005development, shuhua2012city,van2012short,badhrudeen2016short,pan2016short}. Grey models (GM(1,1) and Grey Verhulst) were shown very effective in short-term traffic parameter prediction in \cite{bezuglov2016short, comert2020improved}.  

General short-term traffic prediction is one of the well-investigated, yet still evolving area in transportation with growing demand from connected and autonomous systems. For detailed reviews, the state-of-the-art on short-term forecasting models, and their critical aspects, the reader is referred to works in~\cite{smith2002comparison, van2012short, vlahogianni2014short}. One of the recent review in \cite{vlahogianni2014short} mentions current challenges as combination of existing models, longer prediction horizons with the help of large historical datasets, impact of spatial correlations and transferability, simplicity, adaptability to different data types, handling missing data, continuous or discrete data, and different aggregation levels. Consistent with these points, freeway speed and travel time predictions have been indeed switching towards addressing such issues. Literature review on relevant short-term prediction methods between 2013-2016 can be found in \cite{bezuglov2016short}. Recently, Long-Short-term Memory Neural Networks (LSTM) deep learning approach is applied to flow prediction over 5-min 16 weeks of the Caltrans Performance Measurement System (PeMS). The study aims to detect patterns and reports to improve about $10\%$ \cite{dai2017deeptrend} over Autoregressive Integrated Moving Average (ARIMA) models and about $4\%$ over regular LSTM. In another study, spatial deep learning models are developed for speed prediction on two months of PeMS data from about thousand sensors on multiple weekdays. The results are shown for 15-min or longer predictions with better accuracy $11\%$ and $18\%$ over historical averages (HA) in mean absolute errors (MAE) and MAPE respectively. This is typical with multi-lag ahead prediction results. Modeling trade-off gets pretty narrow especially considering approximately between $5$ mins to $5.4$ hours training time for different versions of deep learning models \cite{cheng2018deeptransport,du2018hybrid,Yu2018Spat}. Mixtures of deep learning methods showed improvements in post accident speed measurements \cite{yu2017deep, li2018diffusion, liao2018deep, deng2017situation}. Another deep learning method of stacked autoencoder (SAE) is used to predict freeway traffic flow from PeMS data. The method is trained with 2 months and tested with a month of flow data. It is compared against BPNN, Random Walk (ARIMA(0,1,0)), SVM, and RBFNN (Radial Basis Function Neural Networks) with aggregated flow data at different levels. The method is consistently better than all compared methods especially for longer-term predictions~\cite{lv2015traffic}. Bayesian networks (BNs) applied for spatial flow prediction indicate error levels for various sensor locations \cite{li2017building}. Similar training computational times are expected for BNs as they need to be trained on large datasets to identify (in)dependencies. Regression tree models tested on 9 weeks of data gives better accuracy than SVM and BPNN methods~\cite{yang2017ensemble}. Travel time is predicted with gradient boosting and extreme gradient boosting giving accurate results over other machine learning methods such as Adaboost and regression trees. Models are able to address complex nonlinear behavior of traffic data and generate travel times for entire network simultaneously. However, model training takes 6.1 and 3.2 hours respectively~\cite{mousa2018comparative}. Gaussian Processes (GPs) with discrete wavelet transforms (i.e., error correction) are given for short-term flow prediction for dynamic network assignment problem~\cite{diao2018hybrid}. GPs are compared against autoregressive, support vector machines (SVM), and BPNN give similar MAE results. This might be due to long-term predictions. Quick and accurate methods are needed for passenger flow prediction in dynamic assignment problems where regression and ARIMA models are applied~\cite{zhang2017real}. Traffic flow prediction with random forest, NNs, regression trees, Nearest Neighbor at work zones is performed using 1-min data for 1,2, and 3-lag ahead predictions. According to the results random forests are the most accurate and regression trees demand the least computational time~\cite{hou2015traffic}. In a k-NN (k-Nearest Neighbors) application, parameters are updated online to better express dynamic traffic behavior and improves the results $3\%$ and $12\%$ over XGB (Extreme Gradientboost) and Seasonal Autoregressive Integrated Moving Average (SARIMA) models respectively~\cite{salamanis2017identifying,duan2018unified,zhang2015component}. 

In summary, prediction models in traffic as well as other fields are switching towards data intensive artificial intelligence models or expert systems. However, concerns with some of these approaches are black box framework, training difficulties, sensitivity analysis, and quick, stochastic and uncontrollable convergence (e.g., meta-heuristics) \cite{van2012difference}. The studies cited above need over a minute training time or minutes in rolling horizon methods. All models tend to rely on big data and computational availability assumptions. Gaussian Processes similar to Grey System theory-based models for traffic parameter prediction can be good candidates especially for real-time system prediction or control implementations with simple interpretable structure, adaptability, and transferability for network applications. Based on these motivations, this paper develops new GPs and applies them to short-term traffic predictions. Developed models are also compared with three Grey models (adopted from \cite{kayacan2010grey}) of first order single variable Grey model (GM(1,1)), GM(1,1) with Fourier error corrections (EFGM), and the Grey Verhulst model with Fourier error corrections (EFGVM). Numerical experiments are presented to demonstrate performance of the GMs against other linear and nonlinear time series models of ARIMA, SARIMA, and Self Exciting Threshold AutoRegressive model (SETAR), Additive Autoregressive model (AAR) \cite{tong1990non,terasvirta1994specification,zivot2007modeling,di2008tsdyn}. For robustness of the method, comparisons are done using four different datasets from California, Virginia, and Portland. 

The rest of the paper is organized as follows. Section~\ref{sctDD} introduces the datasets used in the numerical examples. Section~\ref{sctcomp} focuses on the GP models and covers the Grey systems and time series modes. Section~\ref{sctexp} presents numerical experiments for the compared models and evaluates the traffic prediction performance of the GPs. Finally, section~\ref{sctconc} summarizes the findings and addresses possible future research directions.
 
\section{Data description}
\label{sctDD}
Models are compared in two different datasets obtained from two distinct collection technologies. First dataset is traffic speeds from loop dataset collected by the California PATH (Partners for Advanced Transit and Highways) program on I-$880$ for the Freeway Service Patrol Project. Second dataset is Inrix speed and travel time data from Norfolk and Hampton, Virginia. On the loop data, models are trained and tested with aggregated average traffic speed in miles per hour (mph) observations. 

This dataset (referred as CA) is compiled from $24$ days across $12$~loopsets in February and March of 1993. The dataset contains both normal and abnormal days. Selected loopsets were based on closest to the incident/accident locations. Table~\ref{tab_dataset} summarizes the dataset and assigned ID's to the time series. Each time series consists of two concatenated chains of data, where the first spans from 5:01~AM till 10:00~AM and the second from 2:01~PM till 7:59~PM. The total number of values in each time series is $659$. The gap between the time series (from 10:00AM till 2:01PM) has no significant impact to the models, so each time series is considered to be continuous. 

\begin{figure}[h!]
\centering
\includegraphics[scale=.2]{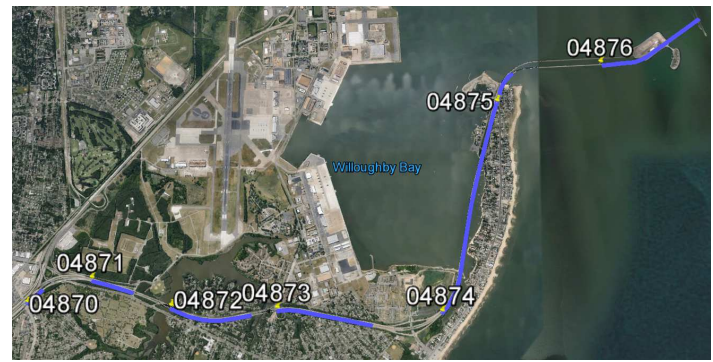}
\caption{Freeway sections used from Inrix data}
\label{fig_inrixfig}       
\end{figure}

INRIX dataset is $20$-second average speed and travel time probe data collected from various freeway sections. The data is aggregated to $1$-minute between $12$:$00$ am-$12$:$00$ pm. At the very  bottom of table~\ref{tab_dataset}, utilized freeway section that is given as $110$+$04874$ with $1.543$ miles (see Fig.~\ref{fig_inrixfig}). The particular section is the longest among all sections available in this study. Since travel times are given in minutes, it is easier to interpret the errors for higher travel times. The part of the dataset used is $23$ weekdays from January $2014$ which are denoted as inserting $I$ in front of digits to distinguish from loop dataset, for instance, $I\#1$ represents January $1$. In addition, Inrix data is collected from freight truck probes' speed and travel times. Whenever no probe is on a link, missing values are repeated. In some cases (e.g., over night hours), long series of values are repeated resulting in flat lines. This causes to generate calculation problems for GM models especially. In order to overcome, very low white noise is added to differentiate values. For speed data $N(0,0.1^2)$ and for travel time data $N(0,0.01^2)$ are added. 
\begin{figure}[h!]
\centering
\includegraphics[scale=.30]{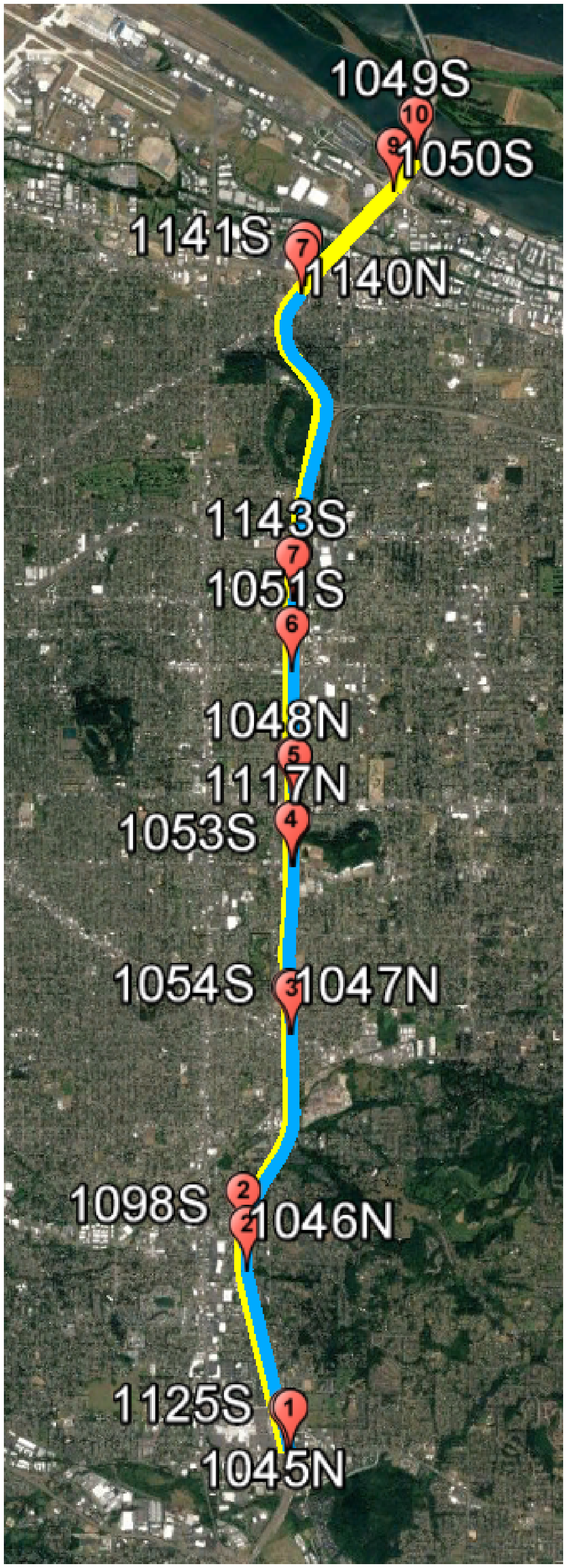}
\caption{Freeway sections used from Portland data}
\label{fig_portlandfig}       
\end{figure}

Last, Portland freeway data consists of two months of 20-second data (September 15, 2011 through November 15, 2011) from dual-loop detectors deployed in the main line and on-ramps. The section of I-205 Northbound (NB) covered by this test data set is 10.09 miles long and the section of I-205 Southbound (SB) covered by this test data set is 12.01 miles long (see Fig.~\ref{fig_portlandfig}). The data samples include flow, occupancy, and speed observations as well as calculated travel times. Data samples from 11 different loop sets are first averaged across all lanes, second processed for missing data which are replaced by interpolations under time series cleaning.

\begin{figure*}[!tbp]
\centering
  \includegraphics[width=0.9\textwidth]{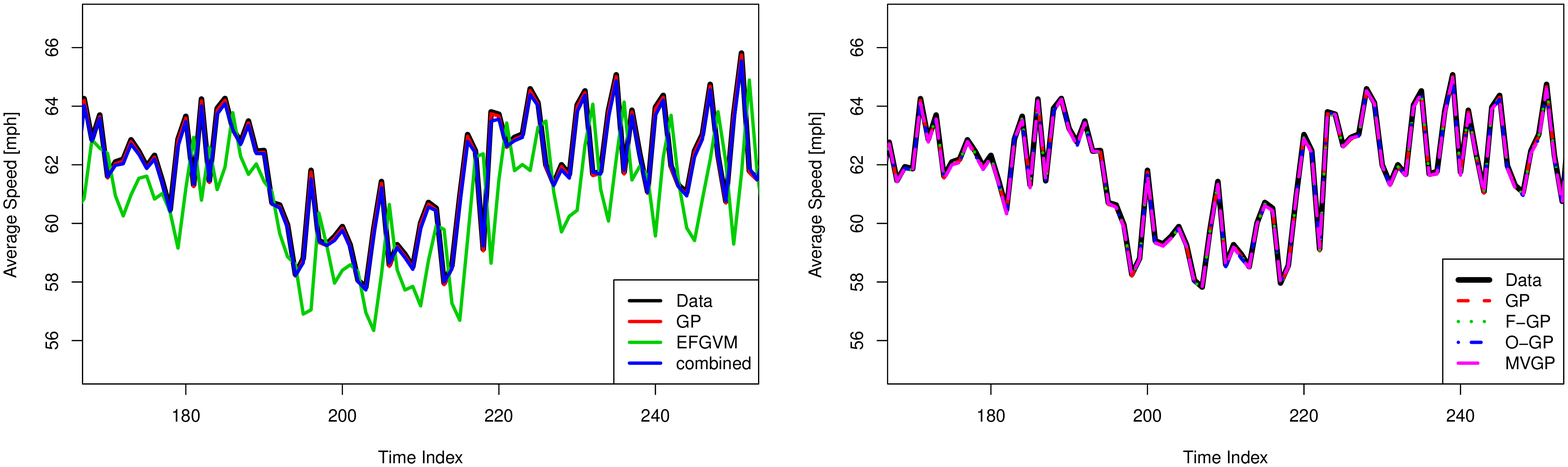}
\caption{Forecasting example on 5-min Portland data}
  \label{fig_fore5}
\end{figure*}

\begin{figure*}[!tbp]
\centering
  \includegraphics[width=0.9\textwidth]{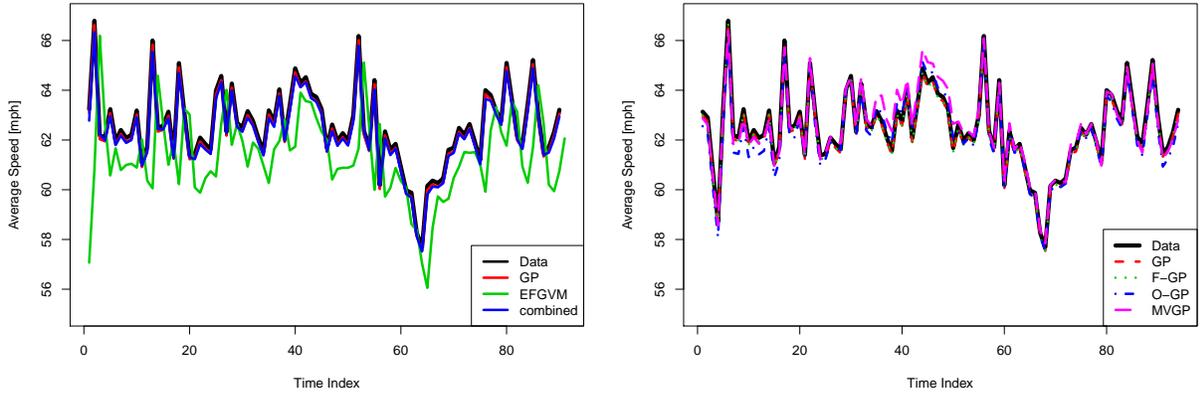}
\caption{Forecasting example on 15-min Portland data}
  \label{fig_fore15}
\end{figure*}

\begin{table}[!htb]
\centering
\caption{Traffic speed and travel time datasets}
\label{tab_dataset} 
\scalebox{0.7}{    
\begin{tabular}{l c c  l c c  l c c}
\hline\noalign{\smallskip}
ID & Date & Loopset & ID & Date & Loopset & ID & Date & Loopset \\
\noalign{\smallskip}\hline\noalign{\smallskip}
\multirow{1}{*}{CA} &  & & \\
\#1 & Feb16 & (5) & \#11 & Mar02 & (17) & \#21 & Mar16 & (12)\\
\#2 & Feb17 & (5) & \#12 & Mar03 & (17) & \#22 & Mar17 & (5)\\
\#3 & Feb18 & (12) & \#13 & Mar04 & (20) & \#23 & Mar18 & (18)\\
\#4 & Feb19 & (19) & \#14 & Mar05 & (20) & \#24 & Mar19 & (19)\\
\#5 & Feb22 & (12) & \#15 & Mar08 & (3) & \\
\#6 & Feb23 & (2) & \#16 & Mar09 & (16) & \\
\#7 & Feb24 & (7) & \#17 & Mar10 & (1) & \\
\#8 & Feb25 & (13) & \#18 & Mar11 & (16) & \\
\#9 & Feb26 & (12) & \#19 & Mar12 & (19) & \\
\#10 & Mar01 & (7) & \#20 & Mar15 & (18)\\
\noalign{\smallskip}\hline\noalign{\smallskip}
\multirow{1}{*}{Inrix} & \#I1&-\#I23  & Jan$1-31$,& $2014$&  $04874$ & \\
\noalign{\smallskip}\hline\noalign{\smallskip}
\multirow{1}{*}{Portland} &13 loops& over &September, &October, &November & 15 & days&\\
\noalign{\smallskip}\hline\noalign{\smallskip}
\end{tabular}
}
\end{table}
Figs.~\ref{fig_fore5}, \ref{fig_fore15}, \ref{fig_port}, and \ref{fig_cal} show average traffic speed, flow, and occupancy time series examples from Portland and CA datasets. Speed data with change points are especially given in Figs.~\ref{fig_port} and \ref{fig_cal} where accurate predictions are more challenging. 

\section{Compared Models}
\label{sctcomp}
In this section, forecasting models are presented. The section contains training of the nonlinear time series models. GPs and GMs can be considered nonparametric models. GPs use kernel or covariance function structure and GMs have time dependent changes with hyperparameters. GMs' hyperparameters can easily be updated in rolling horizon.

\subsection{Gaussian Processes}
\label{sctgp}
Gaussian Processes as natural alternatives with correlation kernel possibly dealing with spatial and temporal correlations as well as collinearities. Gaussian process can be described as regression on function space where over infinitely many dimensions a data generation process is explained via mean $\mu (x)$ and covariance functions $K(x)$ in a semi-parametric way. The only parameters are those in the covariance functions which describe the magnitude and length of the volatility. We applied GP using $tgp$, $GPfit$, $GPFDA$, and $mlegp$ packages in $R$ from \cite{rasmussen2003gaussian, gramacy2007tgp, dancik2008mlegp, shi2014gpfda, macdonald2015gpfit}, respectively. 

For GPs, univariate version uses 1-lagged speed observations, bivariate versions are using flow or occupancy, and multivariate version uses lagged speed, flow, and occupancy observations. Since GPs have covariance structure, although not explicit as time series models, they will handle collinearities among the variables. 

The GP notations are adopted from \cite{turner2012gaussian, shi2014gpfda, macdonald2015gpfit}. Let $X$=$(X_{i1}, X_{i2}, . . . , X_{id})^T$ denote input observations ($i=1,...,n$, $n$ in the numerical examples) with dimension up to $d=3$ in this paper (i.e., lagged speed, flow, and occupancy). Output vector or average speed values are $Y$=$y({X})$=$(Y_1,...,Y_n)^T$ for $Y_i$=$y(X_i)$. We can express $y(X_i)$=$\mu+z(X_i)$ with $\mu$ is overall intercept and we model $z(X_i)$ with GPs $\mu_{z(x_i)}$, ${\sigma_{z(x_i)}}^2$=$\sigma^2$, and covariance function $Cov(z(x_i),z(x_j))$=$\sigma^2R_{ij}$.  

Probability distribution of functions $p(f|X)$=$\mathcal{N}(\mu,K)$ $K_{ij}$ = $K(x_i, x_j)$ with covariance structures experimented are: (1) power exponential $K(x, x'|v,w)$=$ve^{\frac{-w(x-x')^2}{2}}$ (2) rational quadratic $K(x, x'|s,w,a)$=$(1+sw(x-x')^2)^{-a}$, and (3) linear covariance which is non stationary covariance $K(x, x'|a)$= $\sum_{i}{a_ix_ix'_i}$. $n$ training input and output pairs ($X_n$, $Y_n$), and $t$ test inputs ($X_t$, $Y_t$). GPs minimize negative log marginal likelihood with respect to hyperparameters and noise levels.

After carefully adopting and selecting parameters for our problem, we obtained similar results. Out of several kernels or correlation structures (exponential, linear, and Matern). GP fitting minimizes negative log marginal likelihood with respect to hyperparameters and noise level. From our model fit, hyper parameters were calculated for linear kernels GP: $a_1=0.998, \sigma^2=7.869$, GP with flow $a_1=0.975, a_2=2.312\times10^{-6}, \sigma^2=7.867$, GP with occupancy $a_1=0.991, a_2=0.005, \sigma^2=7.831$, and GP with flow and occupancy $a_1=0.908, a_2=1.051\times10^{-10}, a_2=5.053\times10^{-3}, \sigma^2=7.831$ \cite{shi2014gpfda}.  
\subsection{Time Series Models}
\label{sctCFM}
Following nonlinear time series models are used to compare the GP models: Grey models, logistic smooth transition autoregressive model (LSTAR) Eq.(\ref{eqn_lstar}), self-exciting threshold autoregressive model (SETAR) Eq.(\ref{eqn_setar}), neural network nonlinear autoregressive model (NNETS) Eq.(\ref{eqn_nnts}), and additive nonlinear autoregressive model (AAR) Eq.(\ref{eqn_aar}) which are adopted from (\cite{di2015package}). Linear times series models of basic ARIMA and SARIMA models are also fit for comparison. Two types of tests were conducted to understand transferability of the models. 
\begin{enumerate}
\item All of these models are fit to first data series of each dataset listed on Table~\ref{tab_dataset}. Therefore, remaining data series are used for testing. 
\item Models were only fit to California dataset and tested on different datasets.
\end{enumerate}

\begin{eqnarray}
\resizebox{0.9\columnwidth}{!}{$Z_{t+1}=\begin{cases}(\phi_1+\phi_{10}Z_t+\phi_{11}Z_{t-\delta}+...+\phi_{1L}Z_{t-(L-1)\delta})(1-G(Z_t,\gamma,th))+ \\
(\phi_2+\phi_{20}Z_t+\phi_{21}Z_{t-\delta}+...+\phi_{2H}Z_{t-(H-1)\delta})G(Z_t,\gamma,th)+\epsilon_{t+1}\end{cases}$}
\label{eqn_lstar}
\end{eqnarray}
Where $Z_t$ denotes speed or travel time observation at time $t$ and $G(Z_t,\gamma,th)=[1+e^{-\gamma(Z_{t-1}-th)}]^{-1}$ is logistics transition function.

\begin{equation}
\resizebox{1.0\columnwidth}{!}{$Z_{t+1}=\begin{cases}\phi_1+\phi_{10}Z_t+\phi_{11}Z_{t-\delta}+...+\phi_{1L}Z_{t-(L-1)\delta}+\epsilon_{t+1},X_t\leq th\\\phi_2+\phi_{20}Z_t+\phi_{21}Z_{t-\delta}+...+\phi_{2H}Z_{t-(H-1)\delta}+\epsilon_{t+1}, X_t> th\end{cases}$}
\label{eqn_setar}
\end{equation}
Where $L=1$ to $5$ and $H=1$ to $5$ low and high regimes, $X_t$ is threshold function (the transition variable), $\delta$ delay of the transition variable, $th$ is the threshold value. 

\begin{equation}
\resizebox{1.0\columnwidth}{!}{$Z_{t+1}=\beta_0+\sum_{j=1}^D{\beta_ig(\gamma_{0j}+\sum_{i=1}^m{\gamma_{ij}(Z_{t-(i-1)\delta})})}$}
\label{eqn_nnts}
\end{equation}
Eq. (\ref{eqn_nnts}) shows the NNETS model where $m$ denotes embedding dimension, $D$ is hidden layer of the neural network, and $\beta_i$,$\gamma_{0j}$,$\gamma_{ij}$ represent the weights. 

\begin{equation}
Z_{t+1}=\mu+\sum_{j=0}^{m-1}{s_j(Z_{t-(j)\delta})}
\label{eqn_aar}
\end{equation}
In Eq. \ref{eqn_aar}, $s$ represents nonparametric univariate smoothing functions that depends on $Z_t$s and $\delta$ is the delay parameter. Splines from Gaussian family are fitted in the form of $Z_{t+1}\sim\sum_{i=0}^{m-1}{s(Z_{t},..,Z_{t-j})}$. Different $m$ values are fitted and based on their Akaike and Bayesian information criteria (AIC and BIC) values, best models are selected. Following models are determined on the loop and Inrix speed and travel time data respectively: (1) LSTAR $m=3$, $m=2$, $m=2$, (2) SETAR $m=3$, $m=2$, $m=2$, (3) NNETS  $D=4, m=4$, $D=1$, (4) AAR model $m=3$, $m=2$, $m=1$ with Gaussian family function. In addition, forecasting performances are checked and out of relatively close AIC and BIC values, simpler models of are chosen. Resulting fitted models on loop speed, Inrix speed, and Inrix travel time data are presented in Table~\ref{tab_nlfits}. 

\begin{table}[h!]
\centering
\caption{Parameters for nonlinear models trained on November Portland (loop)}
\label{tab_nlfits}       
\scalebox{0.7}{
\begin{tabular}{l l c}
\hline\noalign{\smallskip}
& Model & Parameters \\
\noalign{\smallskip}\hline\noalign{\smallskip}
\multirow{3}{*}{LSTAR(2,2,2)}& L & $\phi_1=4.724$  $\phi_{10}=1.248$  $\phi_{11}=-0.315$  $\phi_{12}=-0.018$ \\
& H &  $\phi_2=23.908$  $\phi_{20}=-1.030$  $\phi_{21}=0.444$  $\phi_{22}=0.219$ \\
& th & $X_t$=$Z_{t}$, th= 63.54\\ \hline
\multirow{4}{*}{SETAR(2,2,2)}& L & $\phi_1=4.736$  $\phi_{10}=1.249$  $\phi_{11}=-0.317$  $\phi_{12}=-0.018$ \\
& H &  $\phi_2=29.744$  $\phi_{20}=0.218$  $\phi_{21}=0.111$  $\phi_{22}=0.202$ \\
& th & $X_t$=$Z_{t}$, th=63.50 \\ 
& proportions & low regime: 53.17\% High regime: 46.83\%  \\ 
\hline
\multirow{1}{*}{AR(3,0,0)}&  & $\mu=4.787$  $\phi_{1}=1.117$  $\phi_{2}=-0.124$ $\phi_{3}=-0.071$ \\ 
\hline
\multirow{2}{*}{SARIMA(1,0,3)(1,0,2)}&  & $\mu=61.473$  $\phi_{1}=0.889$  $\theta_{1}=0.216$ $\theta_{2}=0.107$ $\theta_{3}=0.079$  \\ 
& S &  $\Phi_{1}=0.048$  $\Theta_{1}=-0.138$ $\Theta_{2}=-0.024$ \\  
\hline
\multirow{1}{*}{ARIMA(0,1,1)}&  & $\theta_1=0.156$ \\ 
\hline
\end{tabular}
}
\end{table}

\subsection{Grey Systems}
\label{sctGM}
The Grey Systems theory was developed by Deng~\cite{ju1982control} and since then it has become a preferred method to study and model systems in which the structure or operation mechanism is not completely known~\cite{Deng1989}. According to the theory, the unknown parameters of the system are represented by discrete or continuous grey numbers encoded by symbol $\otimes$. The theory introduces a number of properties and operations on the grey numbers such as the core of the number~$\hat{\otimes}$, its degree of greyness~$g^{\circ}$, and whitenization of the grey number. The latter operation generally describes the preference of the number towards the range of its possible values~\cite{liu2010grey}.

In order to model time series, the theory suggests a family of Grey models, where the basic one is the first order Grey model in one variable further referred to as GM(1,1). The underlying theory in GM(1,1) is as follows. 

Let $X^{(0)}=(x^{(0)}(1),x^{(0)}(2),...,x^{(0)}(n))$ denote a sequence of non-negative observations of a stochastic process and $X^{(1)}=(x^{(1)}(1),x^{(1)}(2),...,x^{(1)}(n))$ be an accumulation sequence of $X^{(0)}$ computed as in Eq.~(\ref{eq:accumulated_seq}).
\begin{equation}
x^{(1)}(k) = \sum_{i=1}^{k}{x^{(0)}(i)}
\label{eq:accumulated_seq}
\end{equation}
then~(\ref{eq:original_GM}) defines the original form of  GM(1,1).
\begin{equation}
x^{(0)}(k) + ax^{(1)}(k) = b
\label{eq:original_GM}
\end{equation}
Let $Z^{(1)}=(z^{(1)}(2),z^{(1)}(3),...,z^{(1)}(n))$ be a mean sequence of $X^{(1)}$ where

\begin{equation}
z^{(1)}(k) = \frac{z^{(1)}(k-1)+z^{(1)}(k)}{2}, \forall k = 2,3,\cdots,n
\label{eq:Z_1}
\end{equation}
Structure of the model is defined using the differential equation Eq.~(\ref{eq:whitenization_GM}) below and then prediction formulas are derived. The subsections below explain the derivations of each model. For the GM models used in this paper, the same set up as in Eqs.~(\ref{eq:accumulated_seq}-\ref{eq:Z_1}) above is followed.
\subsubsection{GM(1,1) model}
The basic form of GM(1,1) is given by the following equation.
\begin{equation}
x^{(0)}(k) + az^{(1)}(k) = b
\label{eq:basic_GM}
\end{equation}
If $\hat{a}=(a,b)^T$ and 
\begin{eqnarray*}
Y = \left[ 
\begin{array}{c}
x^{(0)}(2) \\
x^{(0)}(3) \\
\vdots \\
x^{(0)}(n) \\
\end{array} 
\right],
B = \left[ 
\begin{array}{cc}
z^{(1)}(2) & 1 \\
z^{(1)}(3) & 1 \\
\vdots & \vdots \\
z^{(1)}(n) & 1 \\
\end{array} 
\right]. 
\label{eq:YandB}
\end{eqnarray*}
then, as in~\cite{liu2006grey}, the least squares estimate of the GM(1,1) model is~$\hat{a}=(B^TB)^{-1}B^TY$ and the whitenization equation of the GM(1,1) model is given by,
\begin{equation}
\frac{dx^{(1)}}{dt} + ax^{(1)}(k) = b
\label{eq:whitenization_GM}
\end{equation}
Suppose that $\hat{x}^{(0)}(k)$ and $\hat{x}^{(1)}(k)$ represent the time response sequence (the forecast) and the accumulated time response sequence of GM(1,1) at time $k$ respectively. Then, the latter can be obtained by solving Eq.~(\ref{eq:whitenization_GM}): 
\begin{equation}
\hat{x}^{(1)}(k+1)=\left(x^{(0)}(1)-\frac{b}{a}\right)e^{-ak}+\frac{b}{a}, k=1,2,...,n
\label{eq:model_x_1_solution}
\end{equation}
From Eq.~(\ref{eq:accumulated_seq}), the restored values of $\hat{x}^{(0)}(k+1)$ are calculated as $\hat{x}^{(1)}(k+1)-\hat{x}^{(1)}(k)$. Thus, using Eq.~\ref{eq:model_x_1_solution}, following prediction equation is obtained.
\begin{equation}
\hat{x}^{(0)}(k+1)=\left(1-e^a\right)\left(x^{(0)}(1)-\frac{b}{a}\right)e^{-ak}, k=1,2,...,n
\label{eq:model_x_0_solution}
\end{equation}
Eq.~(\ref{eq:model_x_0_solution}) gives the method to produce forecasts $\forall k=2,3,...,n$. However, for longer time series, a rolling GM(1,1) is preferred. The rolling model observes a window of a few sequential data points in the series: $x^{(0)}(k+1),x^{(0)}(k+2),...,x^{(0)}(k+w)$, where $w \geq 4$ is the window size. Then, the model forecasts one or more future data points: $\hat{x}^{(0)}(k+w+1), \hat{x}^{(0)}(k+w+2)$. The process is repeated for the next $k$. All of the models presented later are utilized in the same way to produce traffic parameter predictions. 
\subsubsection{The Grey Verhulst model (GVM)}
\label{sctGVM}
The response sequence Eq.~(\ref{eq:model_x_0_solution}) implies that the basic GM(1,1) works the best when the time series demonstrate a steady growth or decline and may not perform well when the data has oscillations or saturated sigmoid sequences. For the latter case, the Grey Verhulst model (GVM) is generally used~\cite{liu2010grey}. The basic form of the GVM is present by Eq.~(\ref{eq:verhulst_model}).
%
%
\begin{equation}
x^{(0)}(k)+az^{(1)}(k)=b\left(z^{(1)}(k)\right)^2
\label{eq:verhulst_model}
\end{equation}
The whitenization equation of GVM is:
\begin{equation}
\frac{dx^{(1)}}{dt} + ax^{(1)} = b\left(x^{(1)}\right)^2
\label{eq:verhulst_whitenization}
\end{equation}

Similar to the GM(1,1), the least squares estimate is applied to find~$\hat{a}=(B^TB)^{-1}B^TY$, where 
\begin{eqnarray*}
Y = \left[ 
\begin{array}{c}
x^{(0)}(2) \\
x^{(0)}(3) \\
\vdots \\
x^{(0)}(n) \\
\end{array} 
\right],
B = \left[ 
\begin{array}{cc}
z^{(1)}(2) & z^{(1)}(2)^2 \\
z^{(1)}(3) & z^{(1)}(3)^2 \\
\vdots & \vdots \\
z^{(1)}(n) & z^{(1)}(n)^2 \\
\end{array} 
\right]. 
\label{eq:YandB_VerhulstModel}
\end{eqnarray*}

The forecasts $\hat{x}^{(0)}(k+1)$ are calculated using Eq.~(\ref{eq:verhulst_x_0_solution}).
\begin{equation}
\resizebox{0.9\columnwidth}{!}{$\hat{x}^{(0)}(k+1)=\frac{ax^{(0)}(1)\left(a-bx^{(0)}(1)\right)}{bx^{(0)}(1)+\left(a-bx^{(0)}(1)\right)e^{a(k-1)}}*\frac{\left(1-e^a\right)e^{a\left(k-2\right)}}{bx^{(0)}(1)+\left(a-bx^{(0)}(1)\right)e^{a(k-2)}}$}
\label{eq:verhulst_x_0_solution}
\end{equation}
%
\subsubsection{Error corrections to Grey models}
\label{scterror}
The accuracy of the Grey models can be improved by a few methods. Suppose that $\epsilon^{(0)}$=$\epsilon^{(0)}(1),...,\epsilon^{(0)}(n)$ is the error sequence of $X^{(0)}$, where $\epsilon^{(0)}(k)$= $x^{(0)}(k)-\hat{x}^{(0)}(k)$. If all errors are positive, then a remnant GM(1,1) model can be built (\cite{liu2010grey}). When the errors can be positive or negative, $\epsilon^{(0)}$ can be expressed using Fourier series (\cite{tan1996residual}) as in Eq.~(\ref{eq:fourier_series}).
%
%
%
%
%
\begin{equation}
\epsilon^{(0)}(k) \cong \frac{1}{2}a_0+\sum_{i=1}^{z}\left[a_i cos\left(\frac{2\pi i}{T}k\right)+b_i sin\left(\frac{2\pi i}{T}k\right) \right]
\label{eq:fourier_series}
\end{equation}
where $k = 2,3,...,n$, $T=n-1$, and $z=\left( \frac{n-1}{2}\right)-1$.

The solution is found via the least squares estimate, presuming that $\epsilon^{(0)} \cong PC$ where C is a vector of coefficients: $C=\left[a_0 a_1 b_1 a_2 ... a_n b_n \right]^T$ and matrix P is:
\newcommand\scalemath[2]{\scalebox{#1}{\mbox{\ensuremath{\displaystyle #2}}}}

\begin{eqnarray*}
\scalemath{0.8}{
P = \left[ 
\begin{array}{c c c c c c}
\frac{1}{2} & cos\left(2\frac{2\pi}{T}\right) & sin\left(2\frac{2\pi}{T}\right) & \dots & cos\left(2\frac{2\pi z}{T}\right) & sin\left(2\frac{2\pi z}{T}\right)\\
\frac{1}{2} & cos\left(3\frac{2\pi}{T}\right) & sin\left(3\frac{2\pi}{T}\right) & \dots & cos\left(3\frac{2\pi z}{T}\right) & sin\left(3\frac{2\pi z}{T}\right)\\
\vdots & \vdots & \vdots & \vdots & \vdots & \vdots\\
\frac{1}{2} & cos\left(n\frac{2\pi}{T}\right) & sin\left(n\frac{2\pi}{T}\right) & \dots & cos\left(n\frac{2\pi z}{T}\right) & sin\left(n\frac{2\pi z}{T}\right)\\
\end{array} 
\right] 
\label{eq:Fourier_P}
}
\end{eqnarray*}
%
%
then $C \cong \left(P^T P \right)^{-1}P^T\epsilon^{(0)}$. As a result, the predicted value of the signal has to be corrected according to Eq.~(\ref{eq:fourier_correction}):
\begin{equation}
\hat{x}_f^{(0)}(k) = \hat{x}^{(0)}(k) + \epsilon^{(0)}(k), k = 2,3,...,n
\label{eq:fourier_correction}
\end{equation}
\begin{figure*}[!tbp]
\centering
\begin{subfigure}{.48\textwidth}
\centering
  \includegraphics[width=0.9\linewidth]{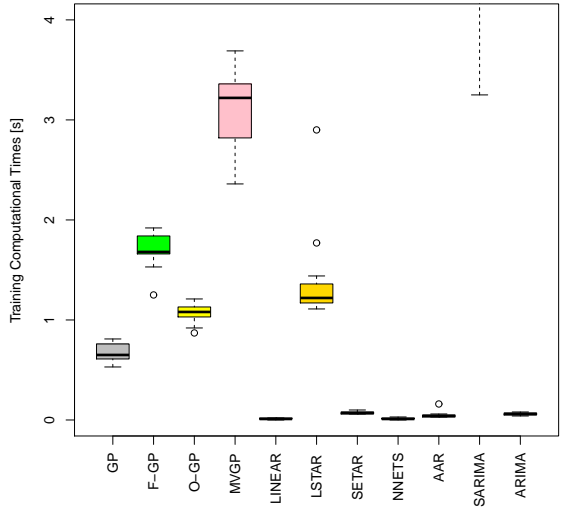}
\caption{Comparison of model training times}
  \label{fig_train}
\end{subfigure}%
\begin{subfigure}{.48\textwidth}
 \centering
\includegraphics[width=0.9\linewidth]{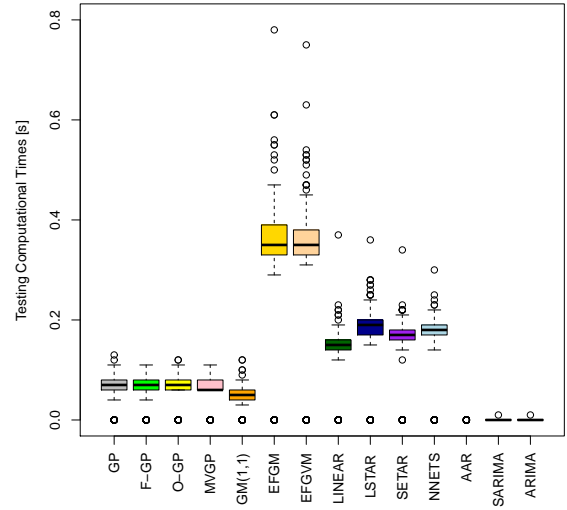}
  \caption{Comparison of model testing times}
  \label{fig_test}
\end{subfigure}
\caption{Errors for $1$-step predictions on 5-min Portland dataset (November)}
\label{fig_cps}
\end{figure*}
\subsection{Computational time}
Computational times from runs are provided in the Fig.~\ref{fig_cps}. Times are obtained in seconds (s) per data series (e.g., Portland 5-min has 288 data values per series on 13 loops over 15 days) on a PC with 8GB of memory, Pentium I5 Quad-Core CPU. GMs do not have training time. Highest training time is required by SARIMA which is beyond the level of other models. GP training times are relatively higher 1-4 seconds and close to LSTAR training times. For testing per time step, models are showing very close results highest with GM(1,1) model. All of the models can be used real time predictions. 
\begin{figure*}[!tbp]
\centering
\begin{subfigure}{.48\textwidth}
\centering
  \includegraphics[width=0.9\linewidth]{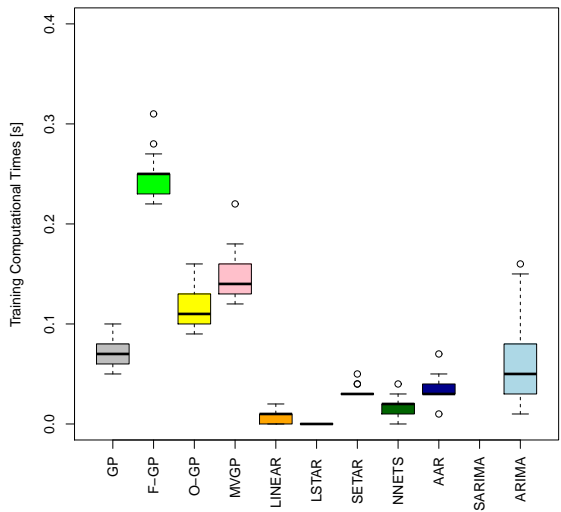}
\caption{Comparison of model training times}
  \label{fig_train2}
\end{subfigure}%
\begin{subfigure}{.48\textwidth}
 \centering
\includegraphics[width=0.9\linewidth]{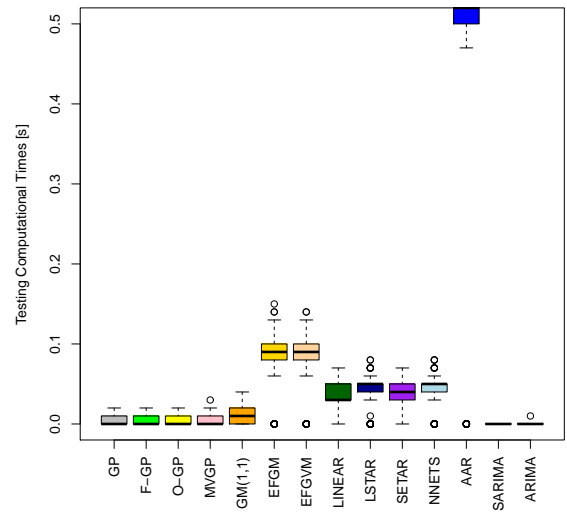}
  \caption{Comparison of model testing times}
  \label{fig_test2}
\end{subfigure}
\caption{Errors for $1$-step predictions on 15-min Portland dataset (November)}
\label{fig_cps2}
\end{figure*}

\subsection{Prediction combination and encompassing}
Prediction combination is used to answer the research question of whether under different conditions (e.g., location and aggregation) different methods are better. A common practice is to combine methods via $f_c=\alpha f_{1}+(1-\alpha) f_{2}$ where $f_1$ and $f_2$ are predictions from different models specifically GPs and GMs. However, as recommended by, this procedure rather for methods using different data sources or information types. In fact, for this study if GPs and GMs are handling predictions rather differently? Intuitively, both methods are handling correlated data which are similar to time series models so combining them should not provide more accurate predictions. Moreover, same data are fed to the models, so, combination does not seem appropriate. Alternatively, this can be checked via forecast encompassing and the proposed model can be further evaluated. Optimum $\alpha^*$ from training or partial testing data can be determined as $\alpha^{*}=(\sum_{t=1}^{T}e_{2t}^2-\sum_{t=1}^{T}e_{1t}e_{2t})/(\sum_{t=1}^{T}e_{1t}^2+\sum_{t=1}^{T}e_{2t}^2-2\sum_{t=1}^{T}e_{1t}e_{2t})$ where $e_{1t}$=$Z_t-\hat{Z}_{1t}$ and $e_{2t}$=$Z_t-\hat{Z}_{2t}$ \cite{clements2008}. 
\begin{figure*}[!tbp]
\centering
\begin{subfigure}{.45\textwidth}
\centering
  \includegraphics[width=0.9\linewidth]{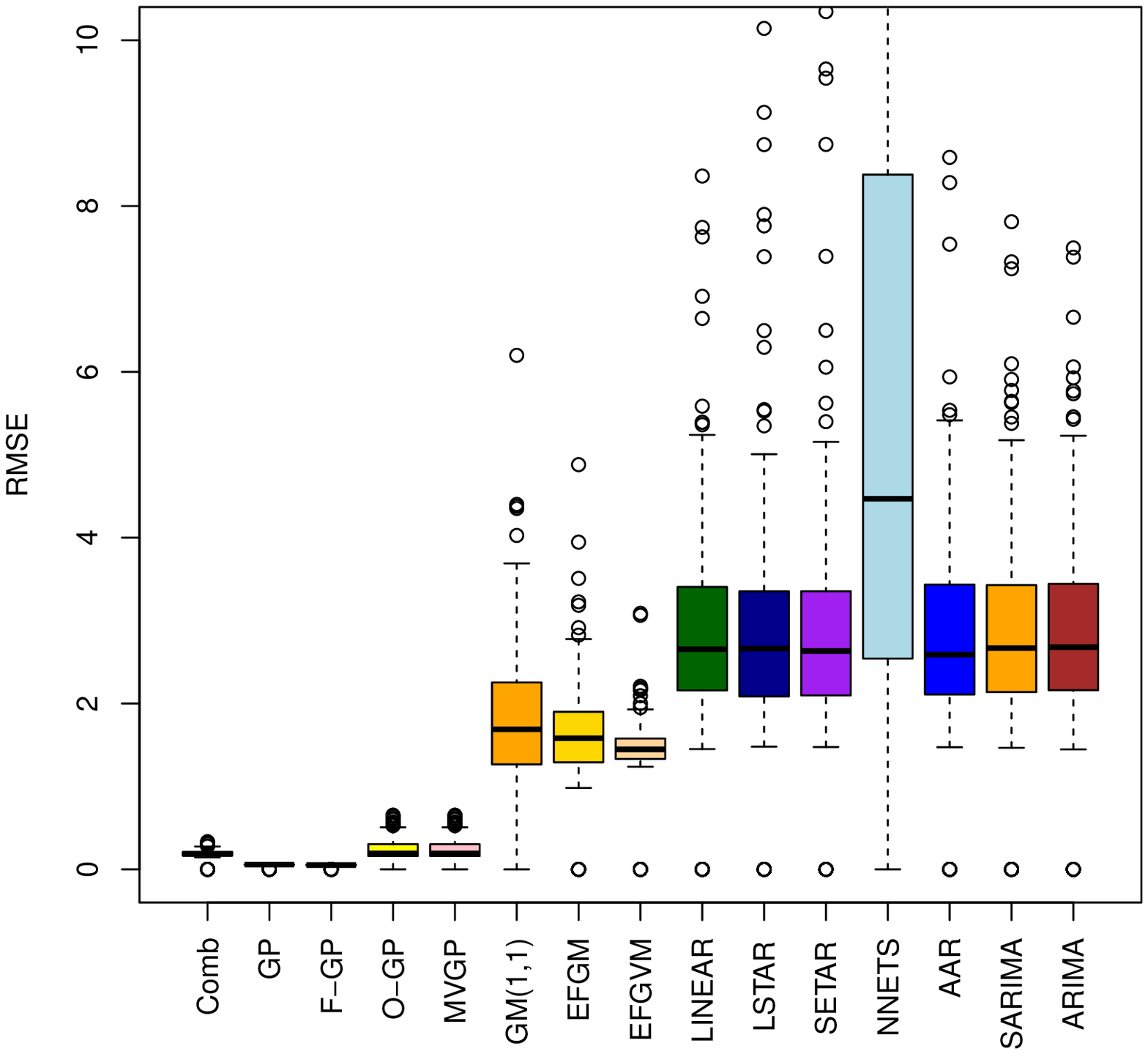}
\caption{Comparison with updated models 5-min Portland}
  \label{fig_en5}
\end{subfigure}%
\begin{subfigure}{.45\textwidth}
 \centering
\includegraphics[width=0.9\linewidth]{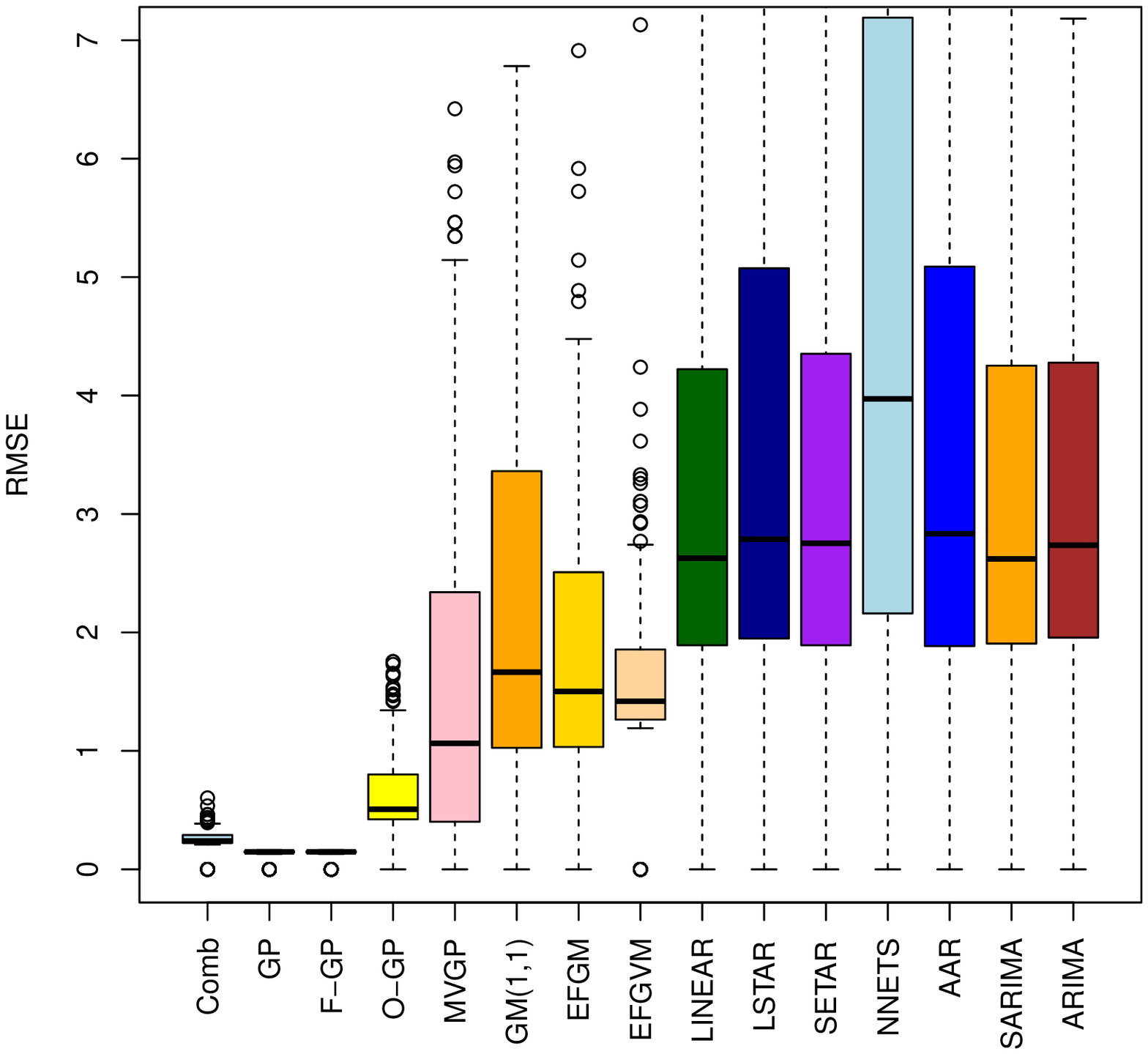}
  \caption{Comparison with updated models 5-min Portland}
  \label{fig_en15}
\end{subfigure}
\caption{Example performance updated models for each dataset for one series}
\label{fig_en}
\end{figure*}

Figs. \ref{fig_fore5} and \ref{fig_fore15} and \ref{fig_en5} and \ref{fig_en15} show the results for combining GP and GM models. It can be seen that there is no contribution of combining GP with GM forecasts. These are obtained from single day results and combination coefficient of $\alpha=0.95$ used in the numerical examples seems reasonable.
\section{Results and Discussion}
\label{sctexp}
\begin{figure*}[!tbp]
\centering
\begin{subfigure}{.45\textwidth}
\centering
  \includegraphics[width=0.95\linewidth]{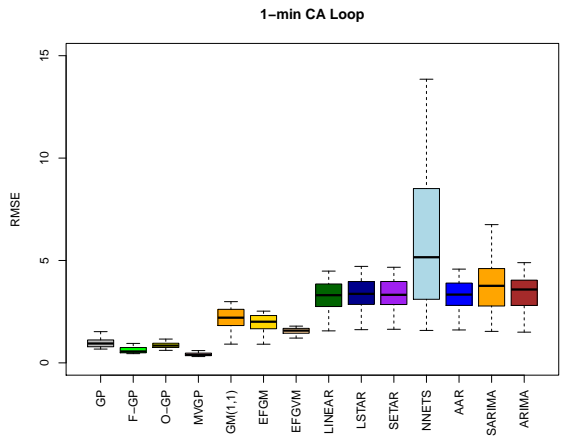}
\caption{CA data 1-min predictions}
  \label{fig_en1}
\end{subfigure}%
\begin{subfigure}{.45\textwidth}
 \centering
\includegraphics[width=0.95\linewidth]{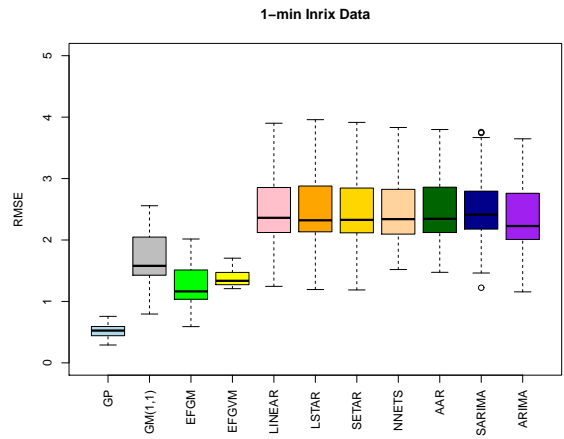}
  \caption{Inrix data 1-min predictions}
  \label{fig_en11}
\end{subfigure}
\caption{Performance of GPs for $1$-step predictions on Inrix and CA data}
\label{fig_cal1}
\end{figure*}

In this section, numerical results are presented for the performance of the applied Gaussian process models and testing of GPs for 1 to 60-minute (min) aggregated 1-step predictions against GM and the nonlinear time series models when models are fit to 1 day California (CA) dataset in order to see performance of the models when transferred to other locations and time. In Figs.~\ref{fig_en1} and \ref{fig_en11}, performance on the highest volatility 1-min data. GPs are able to perform better with multivariate version providing the least errors in root mean squares (RMSE). Inrix dataset do not have flow or occupancy along with speed observations.

\begin{figure*}[!tbp]
\centering
\begin{subfigure}{.45\textwidth}
 \centering
\includegraphics[width=0.95\linewidth]{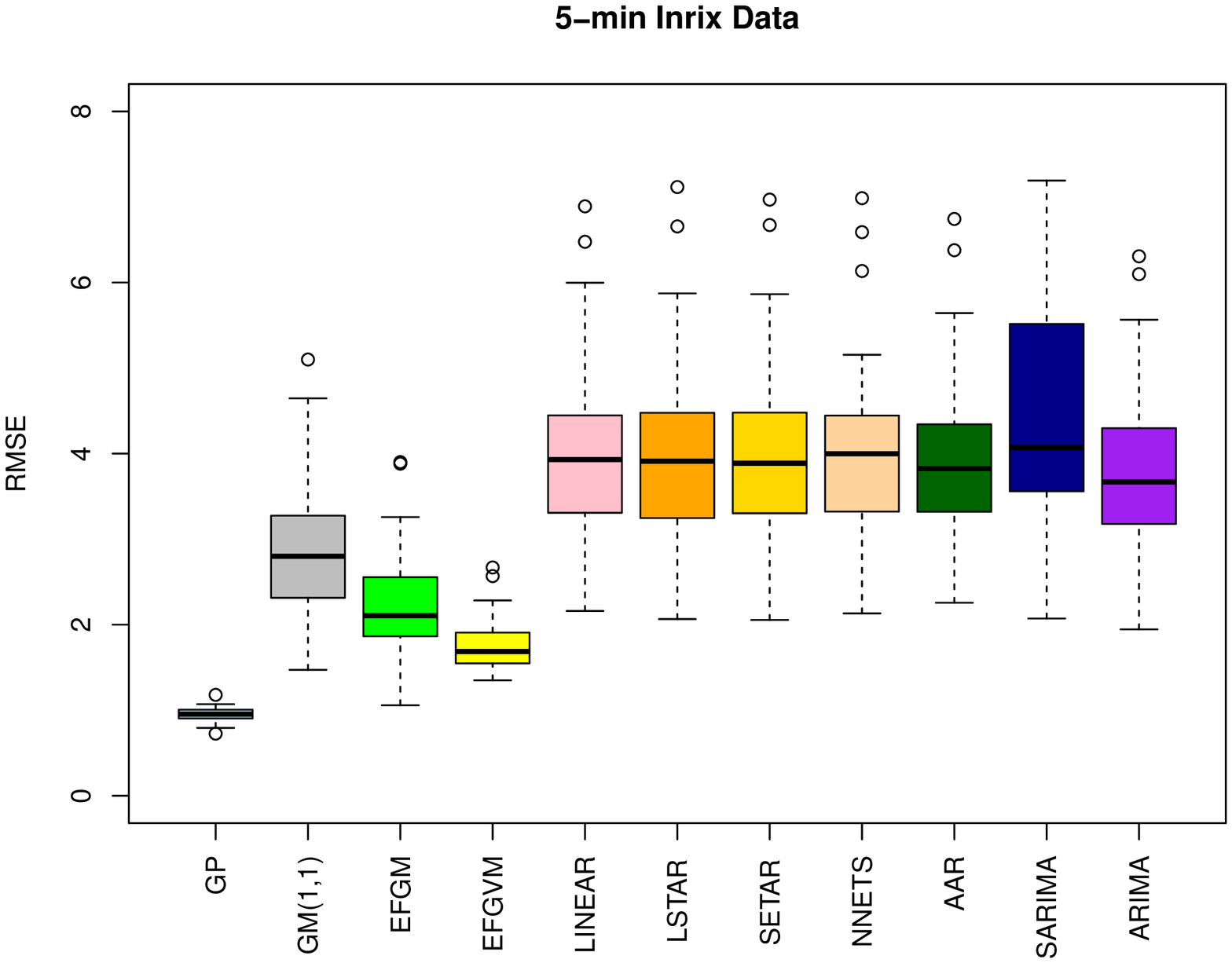}
  \label{fig_inrixgp5}
\end{subfigure}
\begin{subfigure}{.45\textwidth}
\centering
  \includegraphics[width=0.95\linewidth]{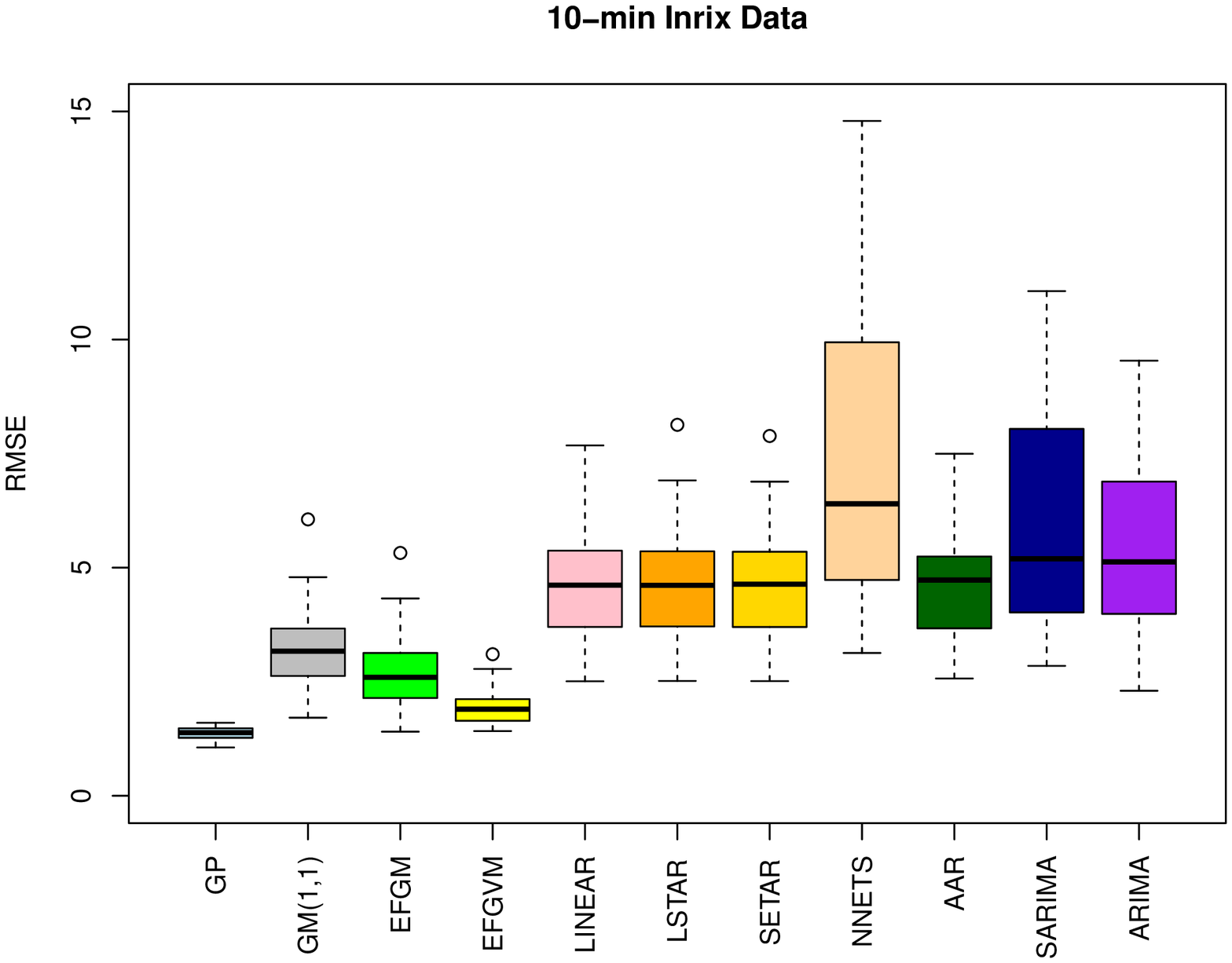}
  \label{fig_inrixgp10}
\end{subfigure}\\
\begin{subfigure}{.45\textwidth}
 \centering
\includegraphics[width=0.95\linewidth]{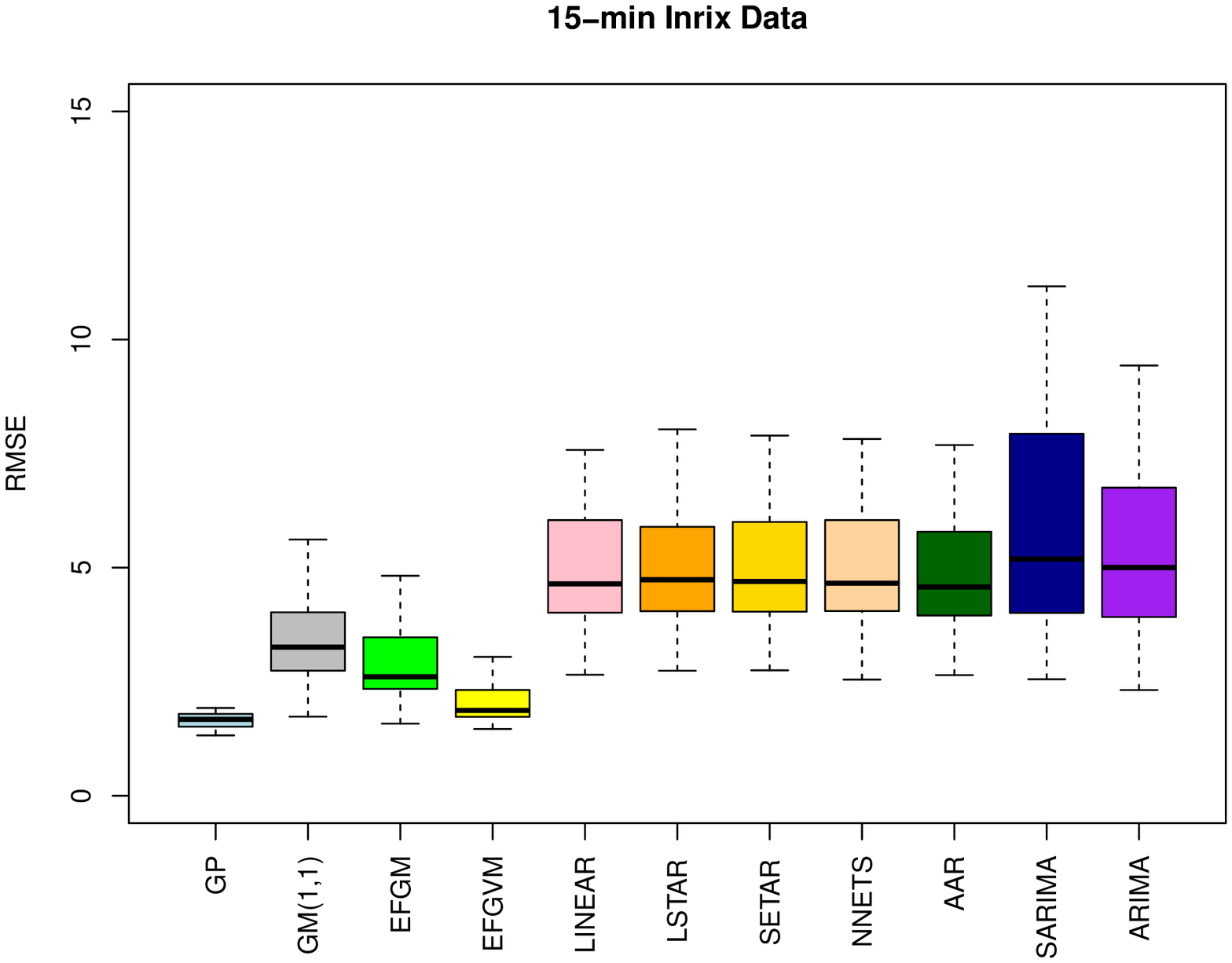}
  \label{fig_inrixgp15}
\end{subfigure}
\begin{subfigure}{.45\textwidth}
 \centering
\includegraphics[width=0.95\linewidth]{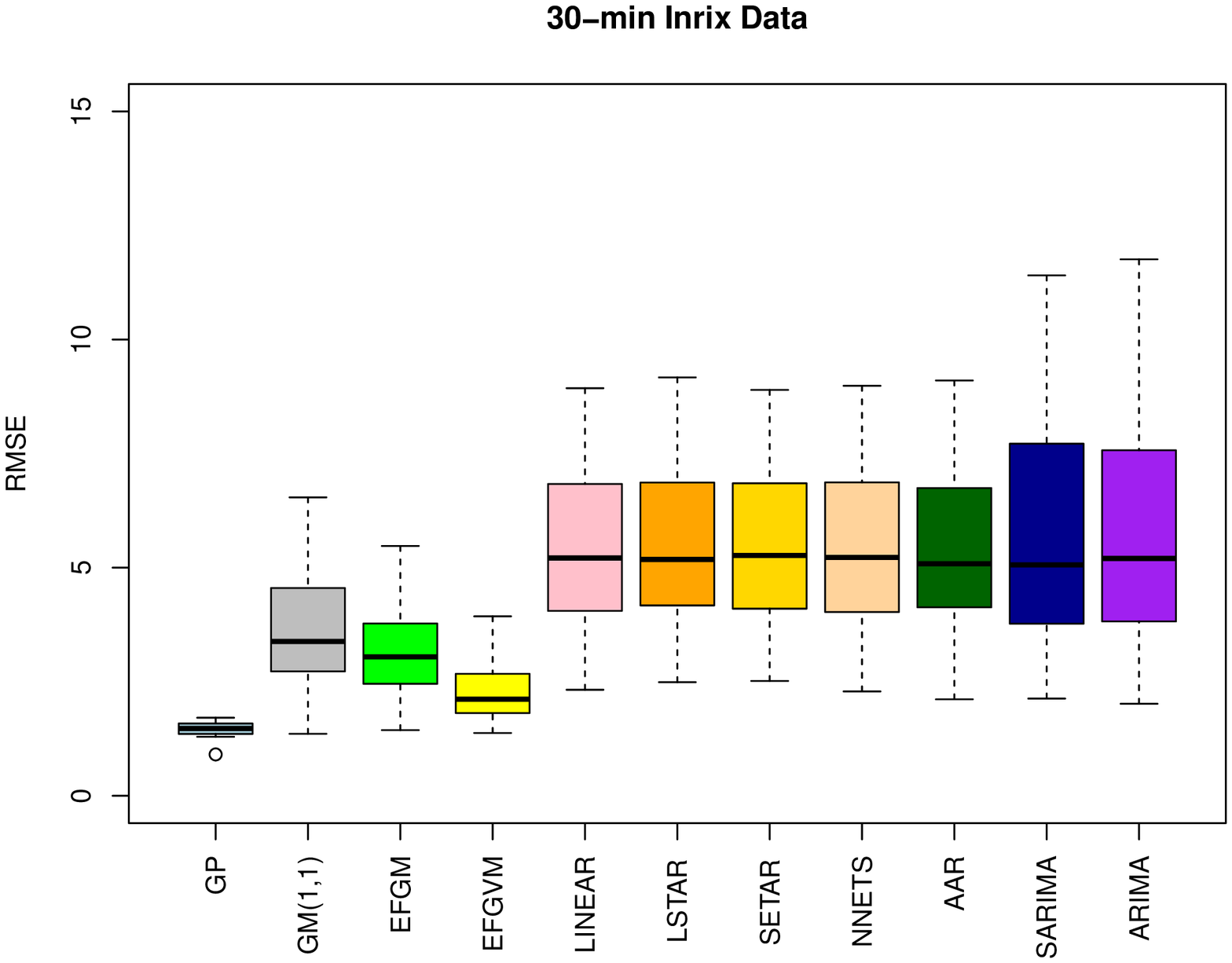}
  \label{fig_inrixgp30}
\end{subfigure}
\begin{subfigure}{.45\textwidth}
 \centering
\includegraphics[width=0.95\linewidth]{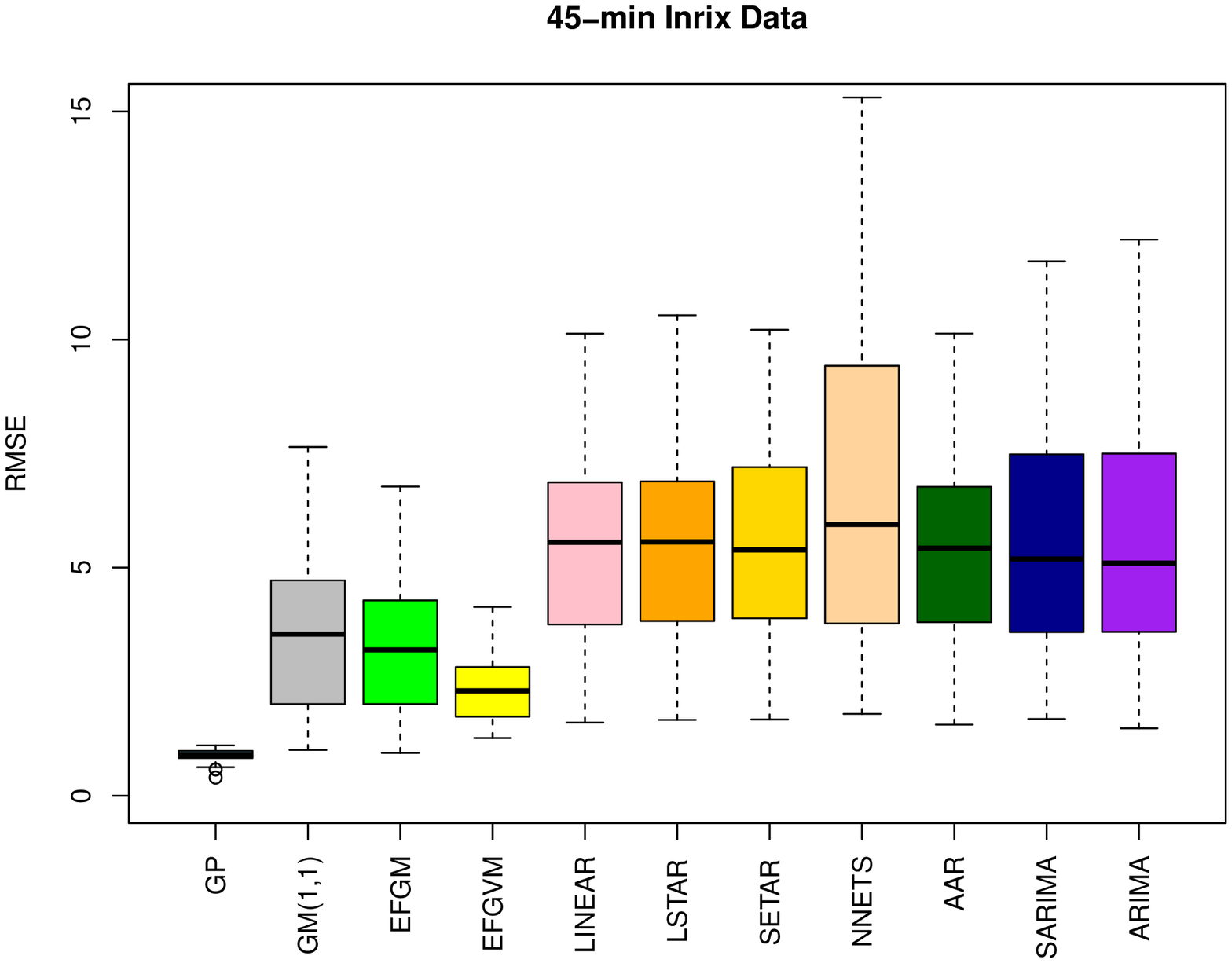}
  \label{fig_inrixgp45}
\end{subfigure}
\begin{subfigure}{.45\textwidth}
 \centering
\includegraphics[width=0.95\linewidth]{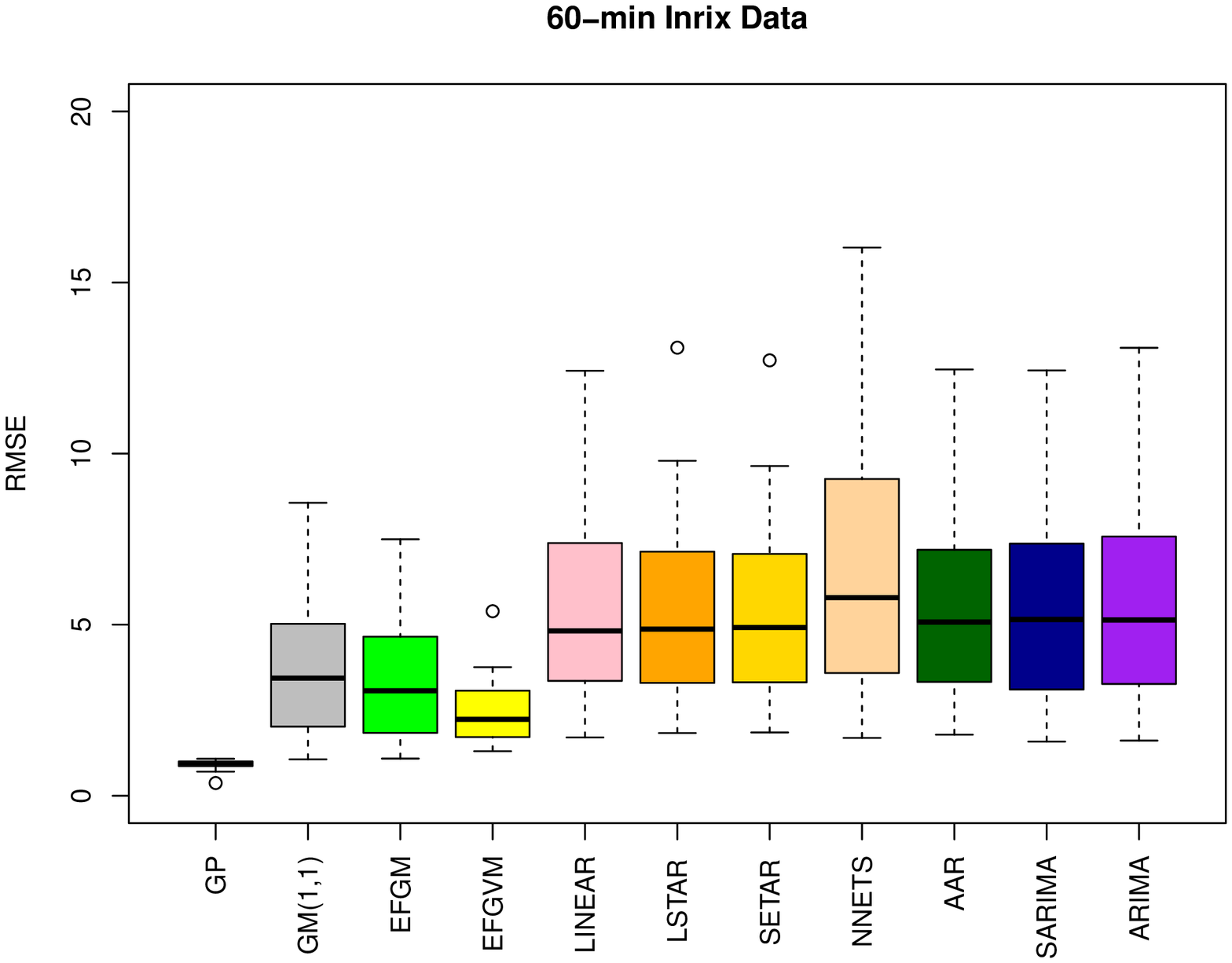}
  \label{fig_inrixgp60}
\end{subfigure}
\caption{Errors for $1$-step predictions on loop and Inrix dataset}
\label{fig_inrixgp}
\end{figure*}

In Fig.~\ref{fig_inrixgp} presents the results for Inrix probe vehicle dataset. In this dataset, there is no flow or occupancy data along with speed observations. Thus, GPs and compared models are compared on different aggregation levels only. Simply, GPs are able to provide 5-$min$ aggregated forecasts within 1 mile per hour ($mph$) root mean squared errors (RMSE) of average speed level of 60 $mph$ (i.e., $1.67 \%$ of average speed level). GPs are able to provide close accuracy levels at all aggregations up to 60 $min$. The results are consistently better than the GM or other compared models.  

\begin{figure*}[!tbp]
\centering
\begin{subfigure}{.45\textwidth}
 \centering
\includegraphics[width=0.95\linewidth]{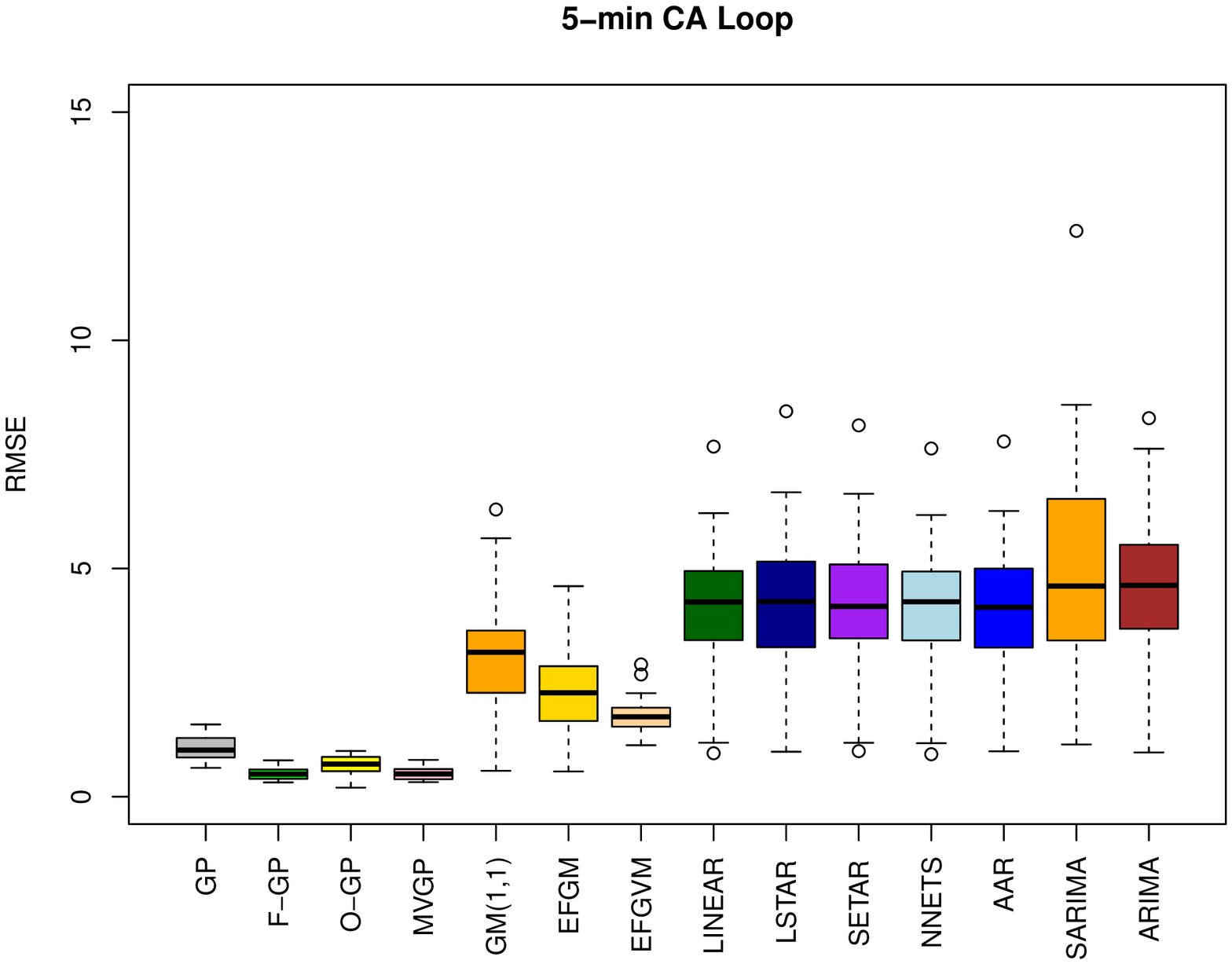}
  \label{fig_calgp5}
\end{subfigure}
\begin{subfigure}{.45\textwidth}
\centering
  \includegraphics[width=0.95\linewidth]{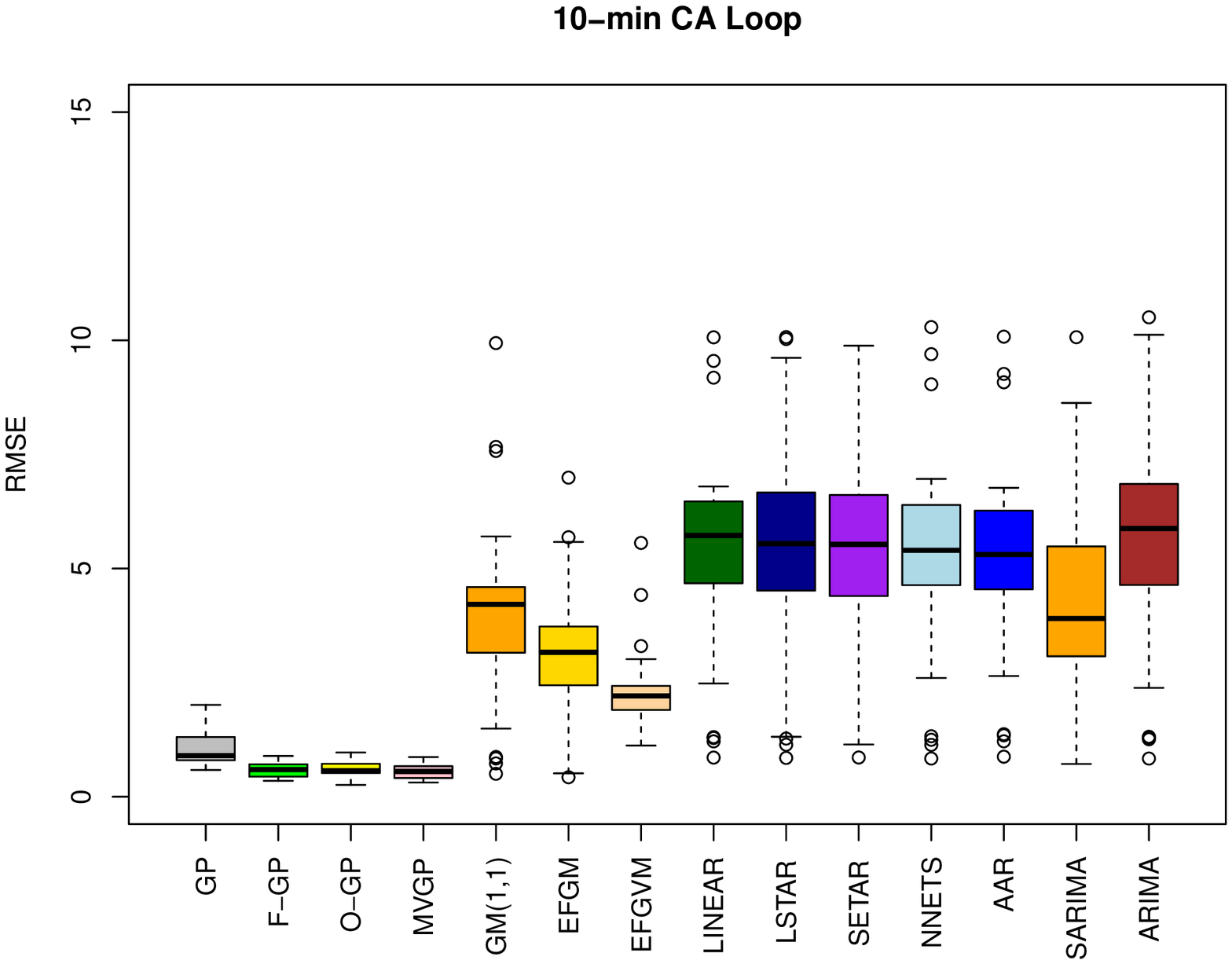}
  \label{fig_calgp10}
\end{subfigure}\\
\begin{subfigure}{.45\textwidth}
 \centering
\includegraphics[width=0.95\linewidth]{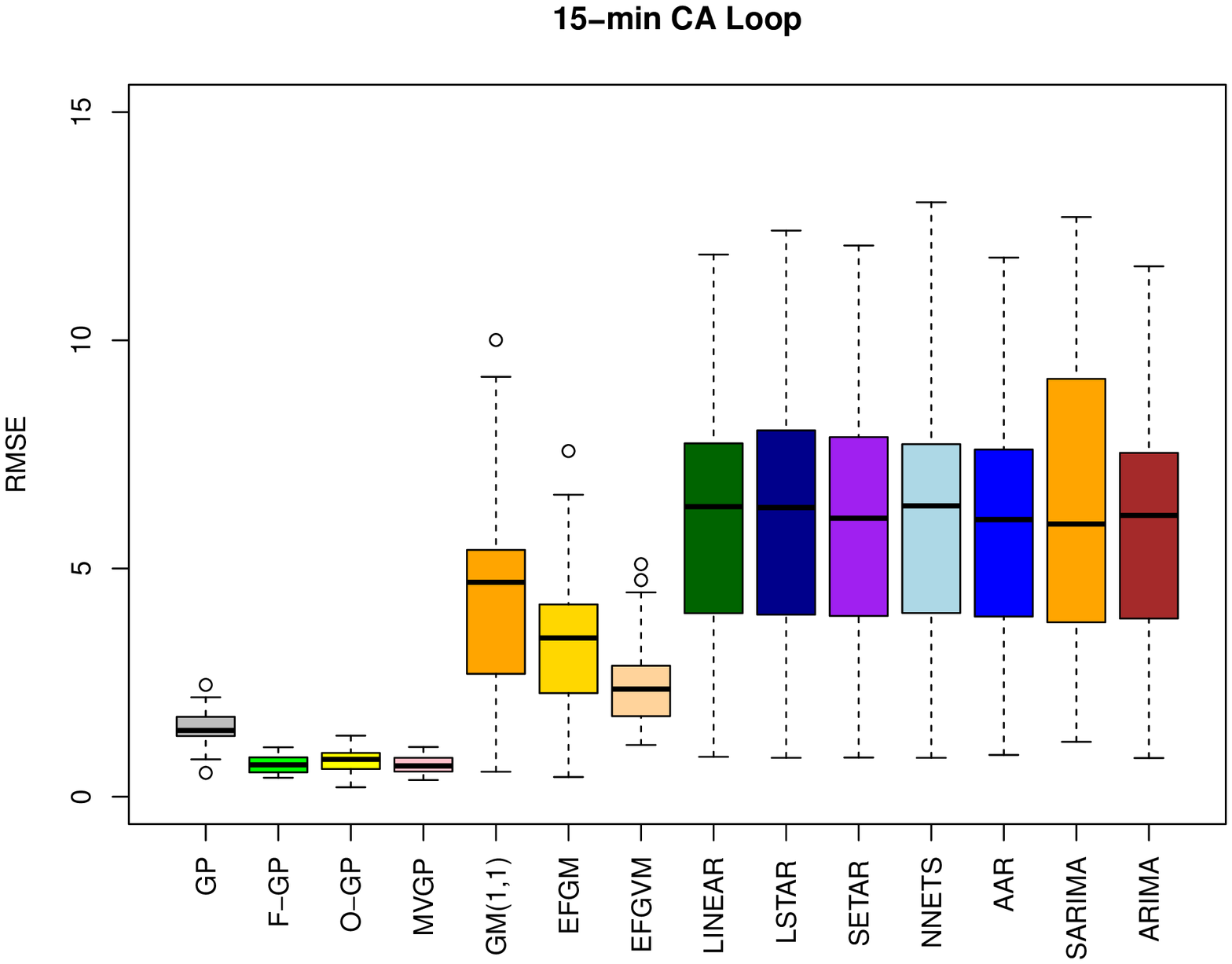}
  \label{fig_calgp15}
\end{subfigure}
\begin{subfigure}{.45\textwidth}
 \centering
\includegraphics[width=0.95\linewidth]{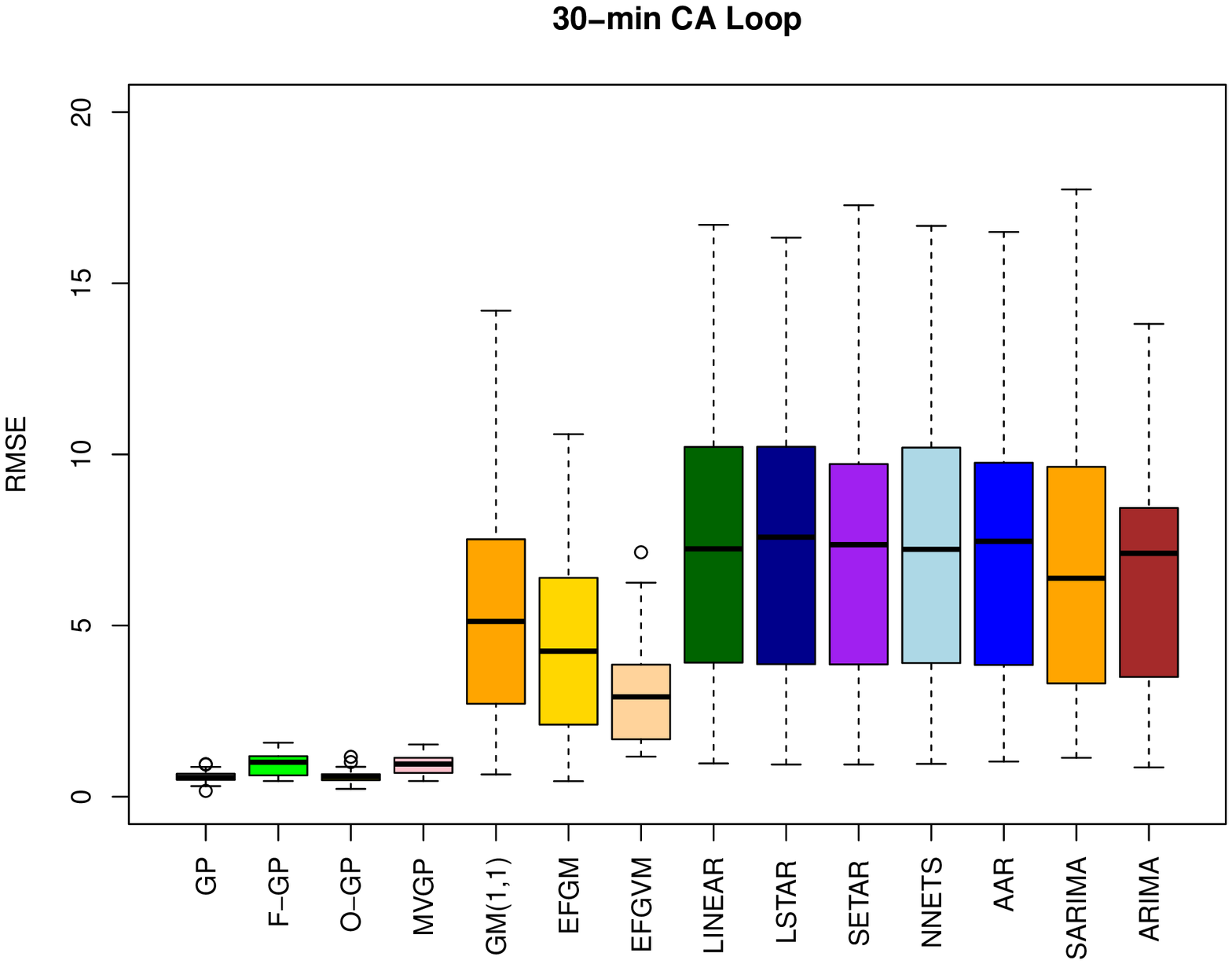}
  \label{fig_calgp30}
\end{subfigure}
\begin{subfigure}{.45\textwidth}
 \centering
\includegraphics[width=0.95\linewidth]{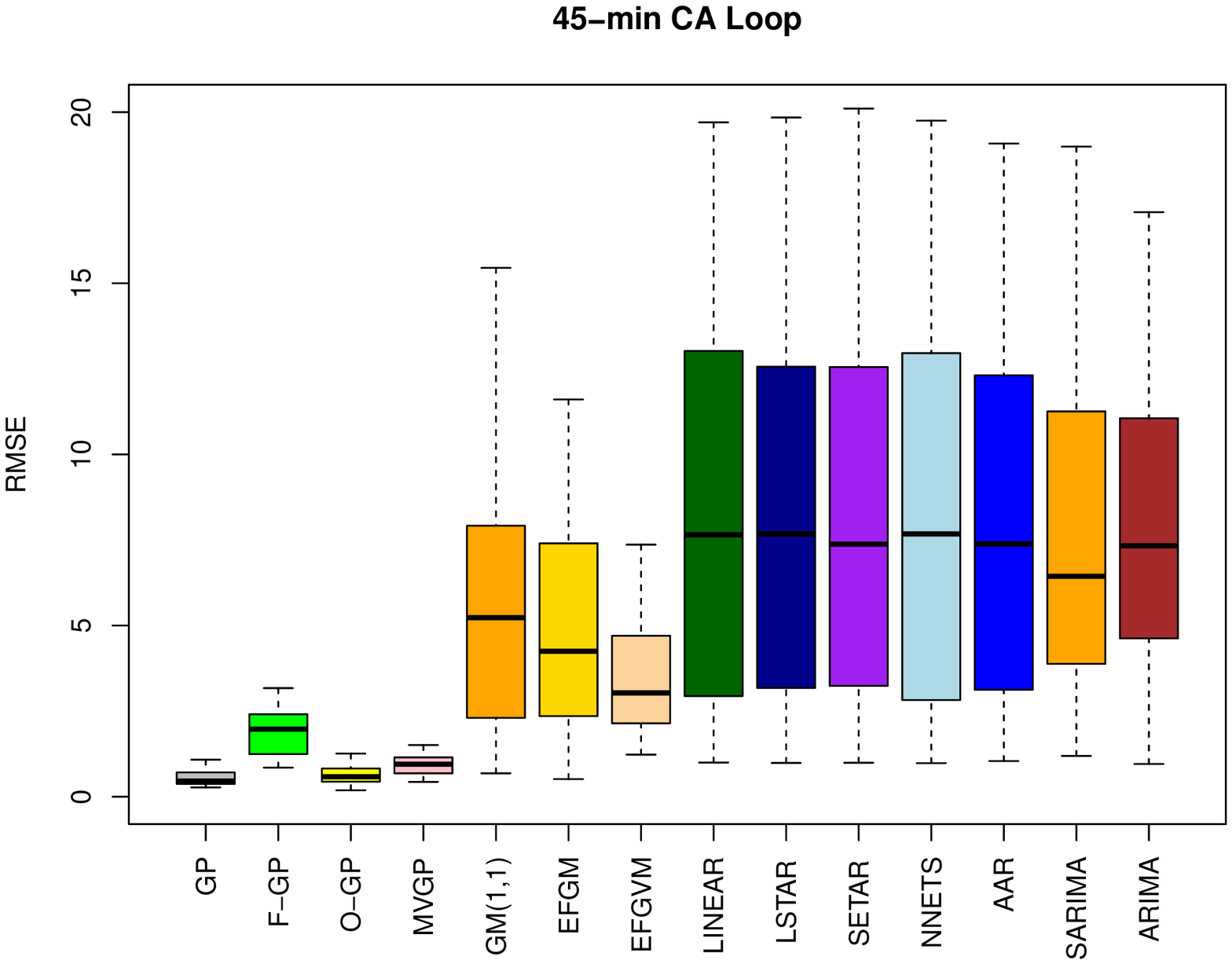}
  \label{fig_calgp45}
\end{subfigure}
\begin{subfigure}{.45\textwidth}
 \centering
\includegraphics[width=0.95\linewidth]{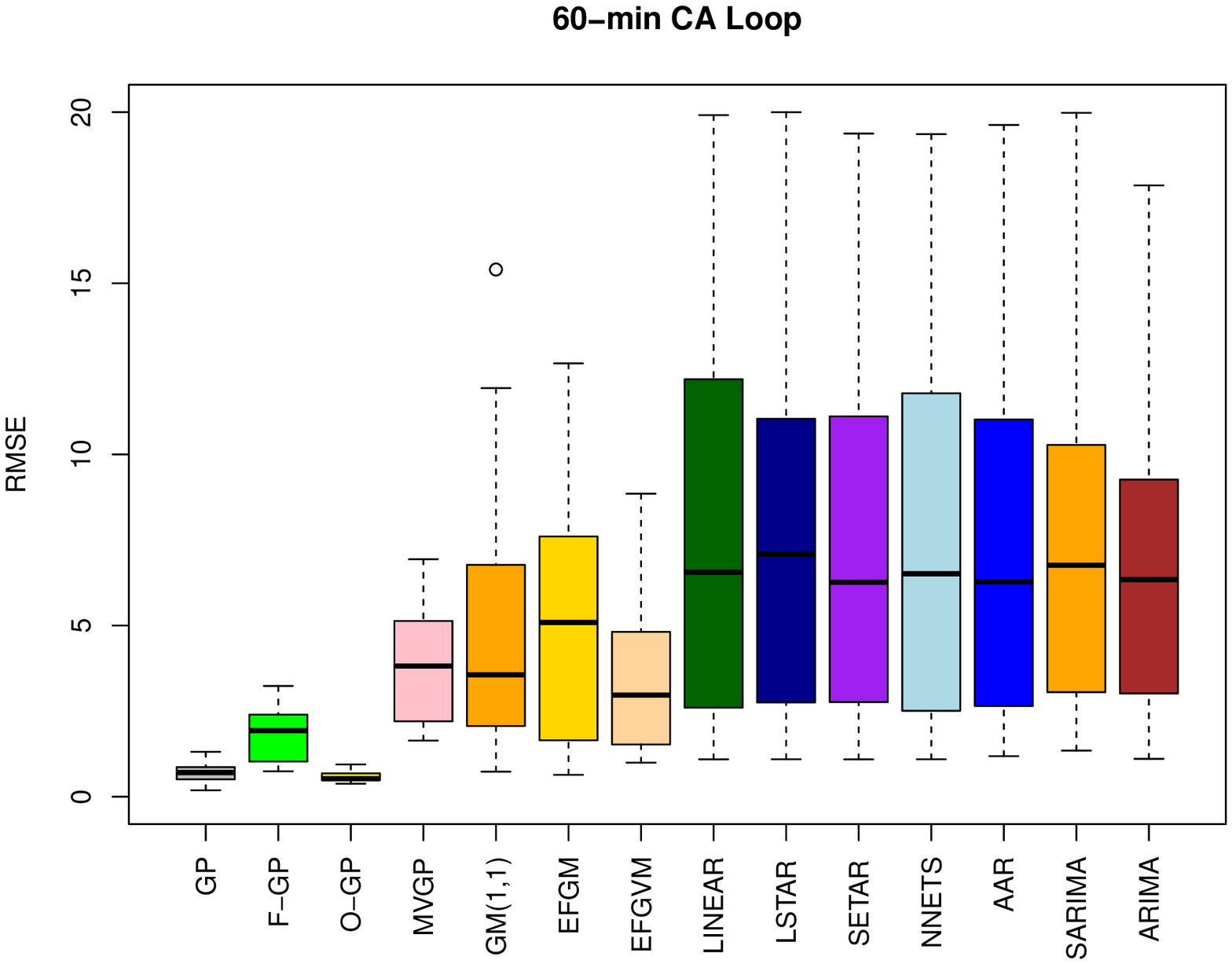}
  \label{fig_calgp60}
\end{subfigure}
\caption{Errors for $1$-step predictions on CA loop dataset}
\label{fig_calgp}
\end{figure*}

In CA dataset Fig.~\ref{fig_calgp} shows up to 15-min aggregation level, bivariate and multivariate versions of GP are better forecasting. After 30-min aggregation levels, impact of multivariate forecasting including flow and occupancy decreases. However, occupancy is able to help speed prediction accuracy consistently. This can be simply explained as speed and density has approximately negatively linearly related (e.g., by Greenshields model flow$=$speed$\times$density). 

\begin{figure*}[!tbp]
\centering
\begin{subfigure}{.45\textwidth}
 \centering
\includegraphics[width=0.95\columnwidth]{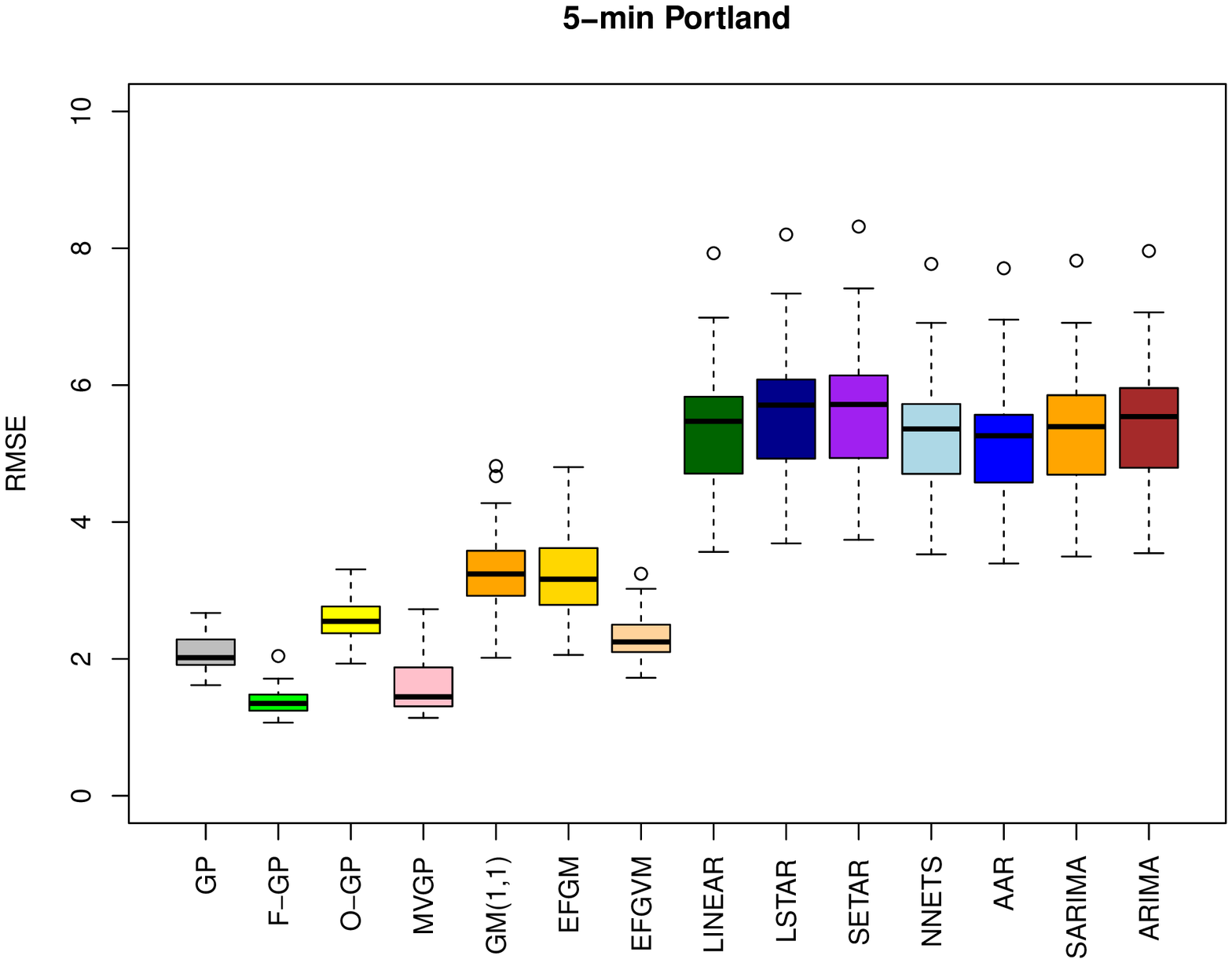}
  \label{fig_portgp5}
\end{subfigure}
\begin{subfigure}{.45\textwidth}
\centering
  \includegraphics[width=0.95\columnwidth]{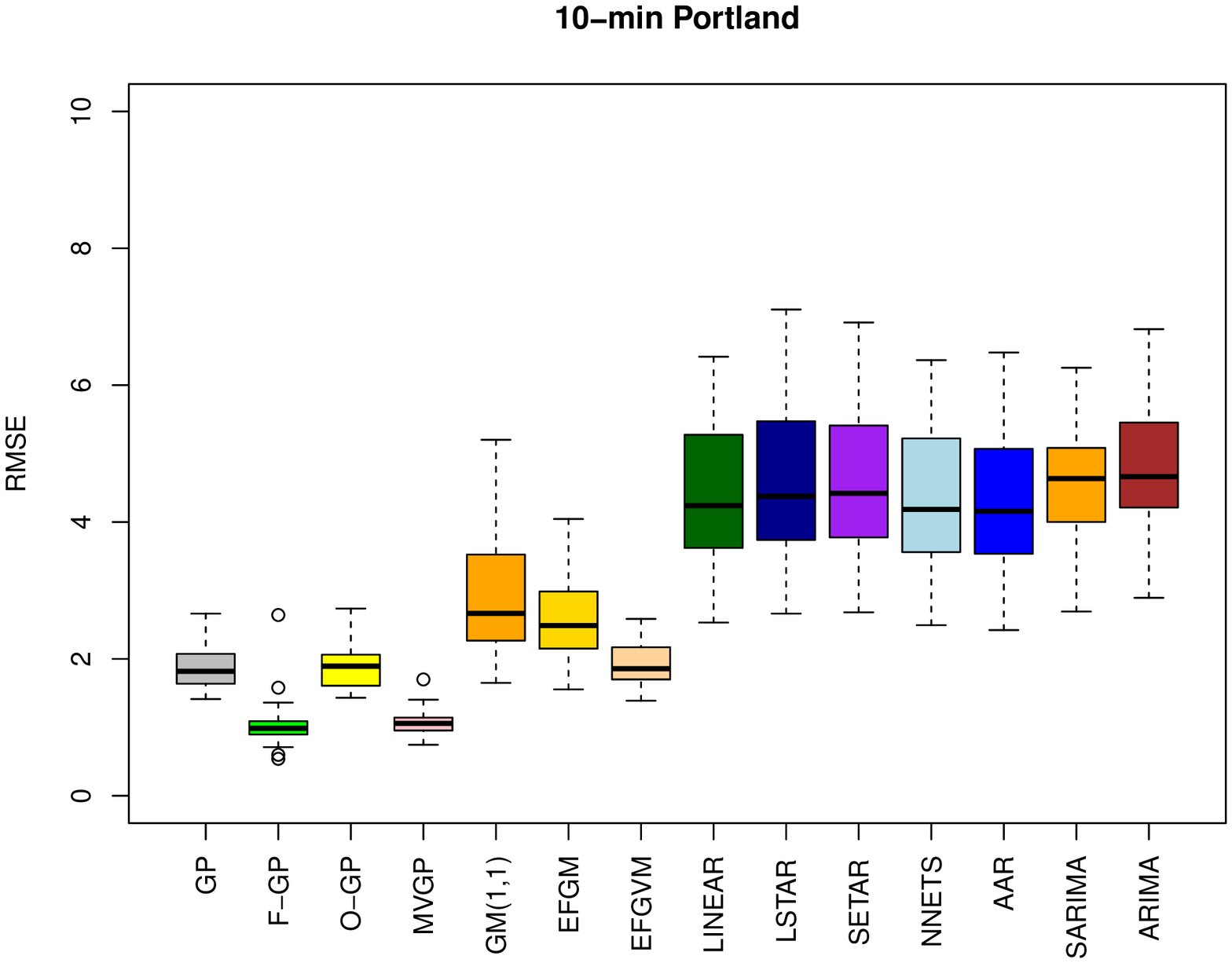}
  \label{fig_portgp10}
\end{subfigure}\\
\begin{subfigure}{.45\textwidth}
 \centering
\includegraphics[width=0.95\columnwidth]{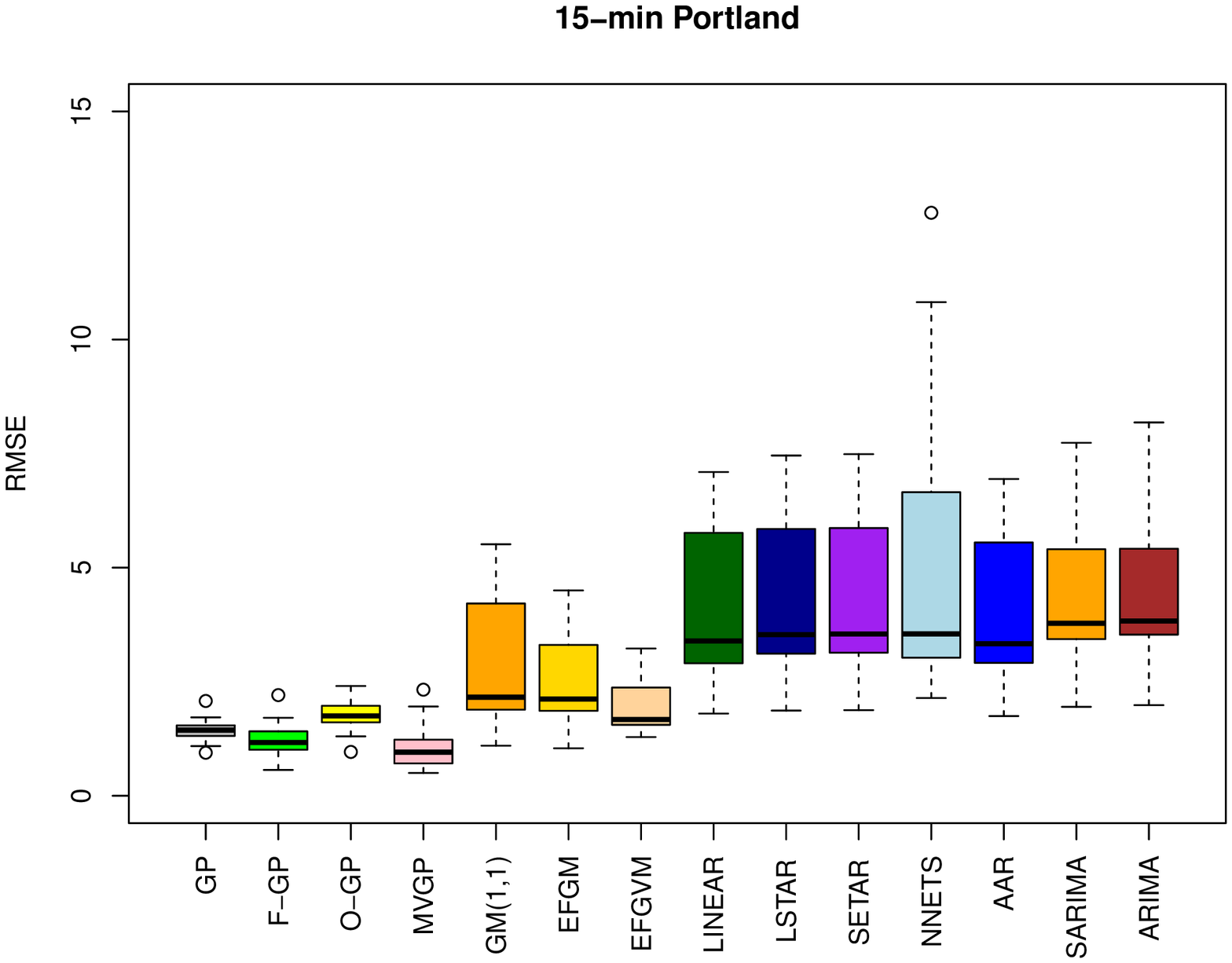}
  \label{fig_portgp15}
\end{subfigure}
\begin{subfigure}{.45\textwidth}
 \centering
\includegraphics[width=0.95\columnwidth]{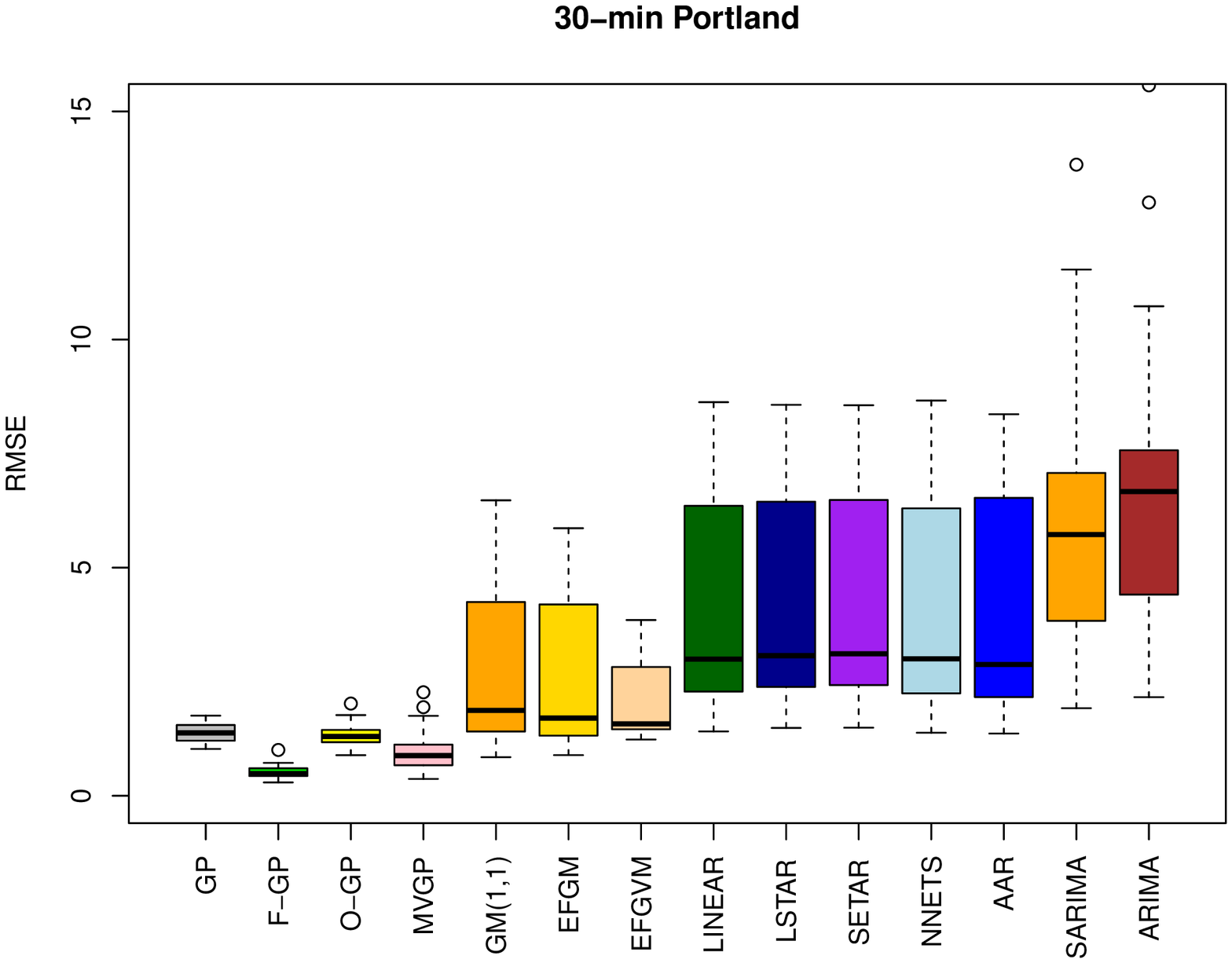}
  \label{fig_portgp30}
\end{subfigure}
\begin{subfigure}{.45\textwidth}
 \centering
\includegraphics[width=0.95\columnwidth]{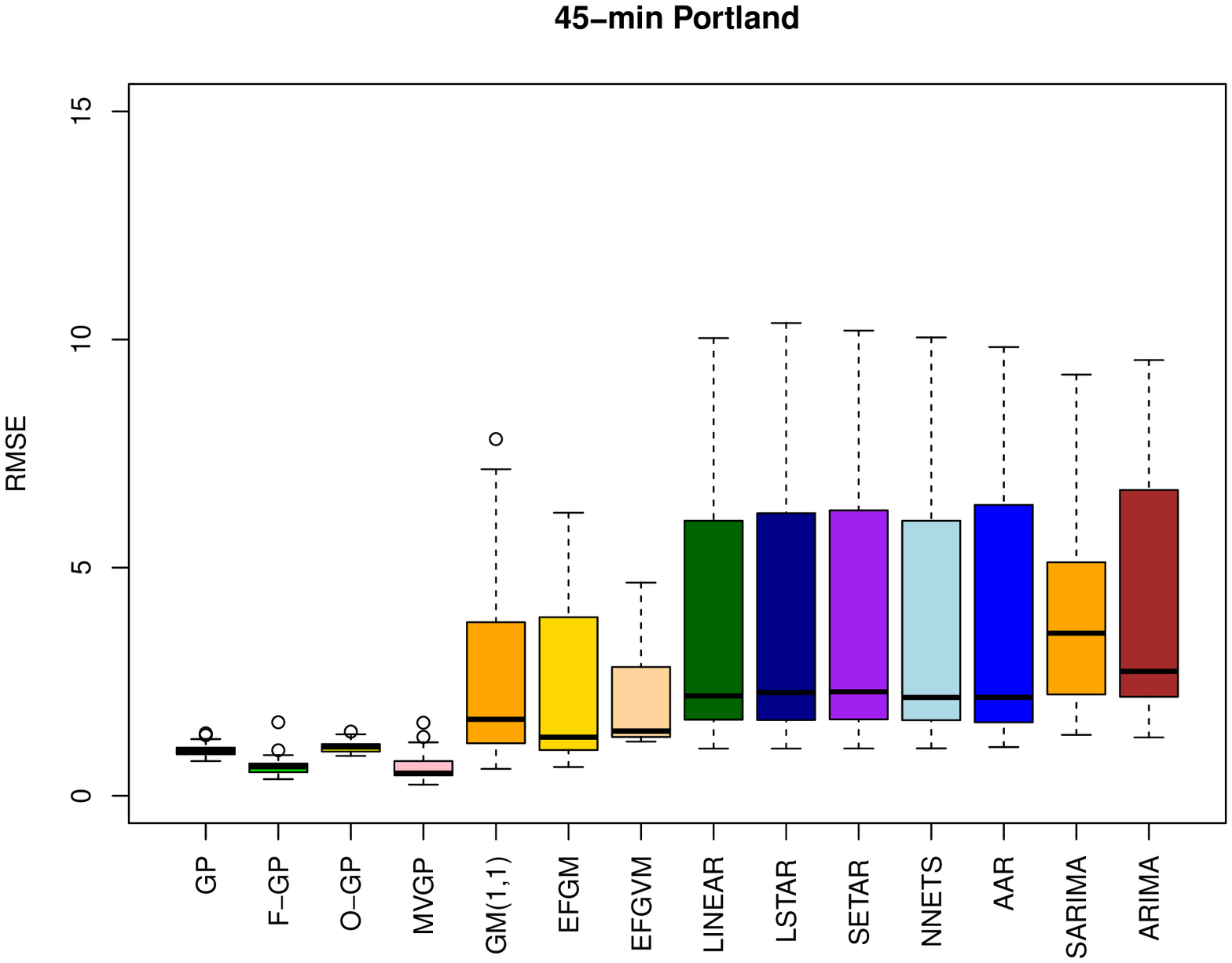}
  \label{fig_portgp45}
\end{subfigure}
\begin{subfigure}{.45\textwidth}
 \centering
\includegraphics[width=0.95\columnwidth]{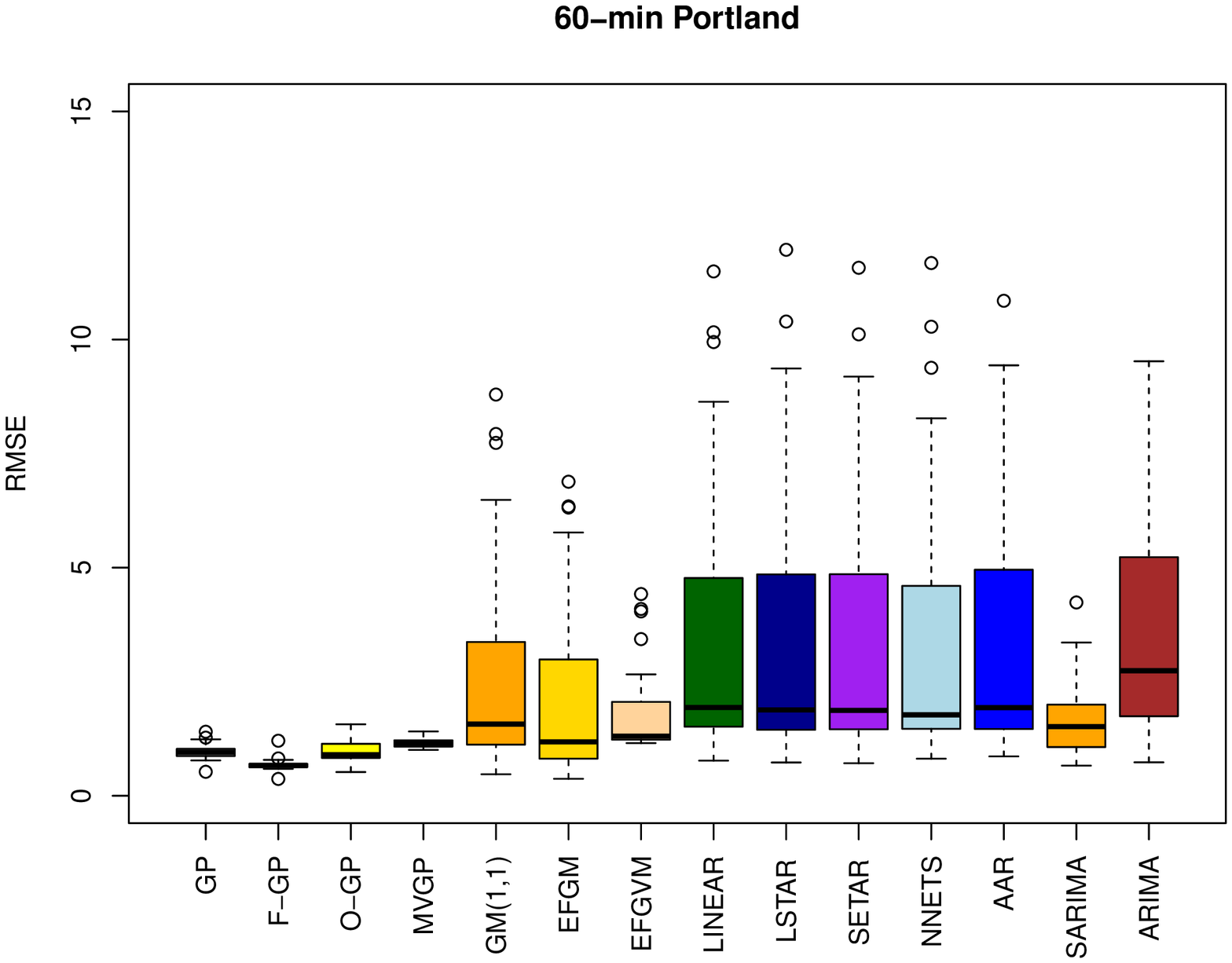}
  \label{fig_portgp60}
\end{subfigure}
\caption{Errors for $1$-step predictions on Portland loop dataset}
\label{fig_portgp}
\end{figure*}

Lastly, the results for the freeway loop dataset referred as Portland dataset are presented in Fig.~\ref{fig_portgp}. For this dataset up to 45-min aggregation levels, bivariate and multivariate GP versions are providing more accurate predictions. In this dataset, better contributing variable is flow as opposed to occupancy. This is due to lack of dynamic series present in overall dataset. In order to challenge the models, CA dataset particularly contains the days with accidents with multilane closures. Portland dataset contains both normal and the days with accidents. 

\begin{figure}[!tbp]
\centering
  \includegraphics[width=0.99\linewidth]{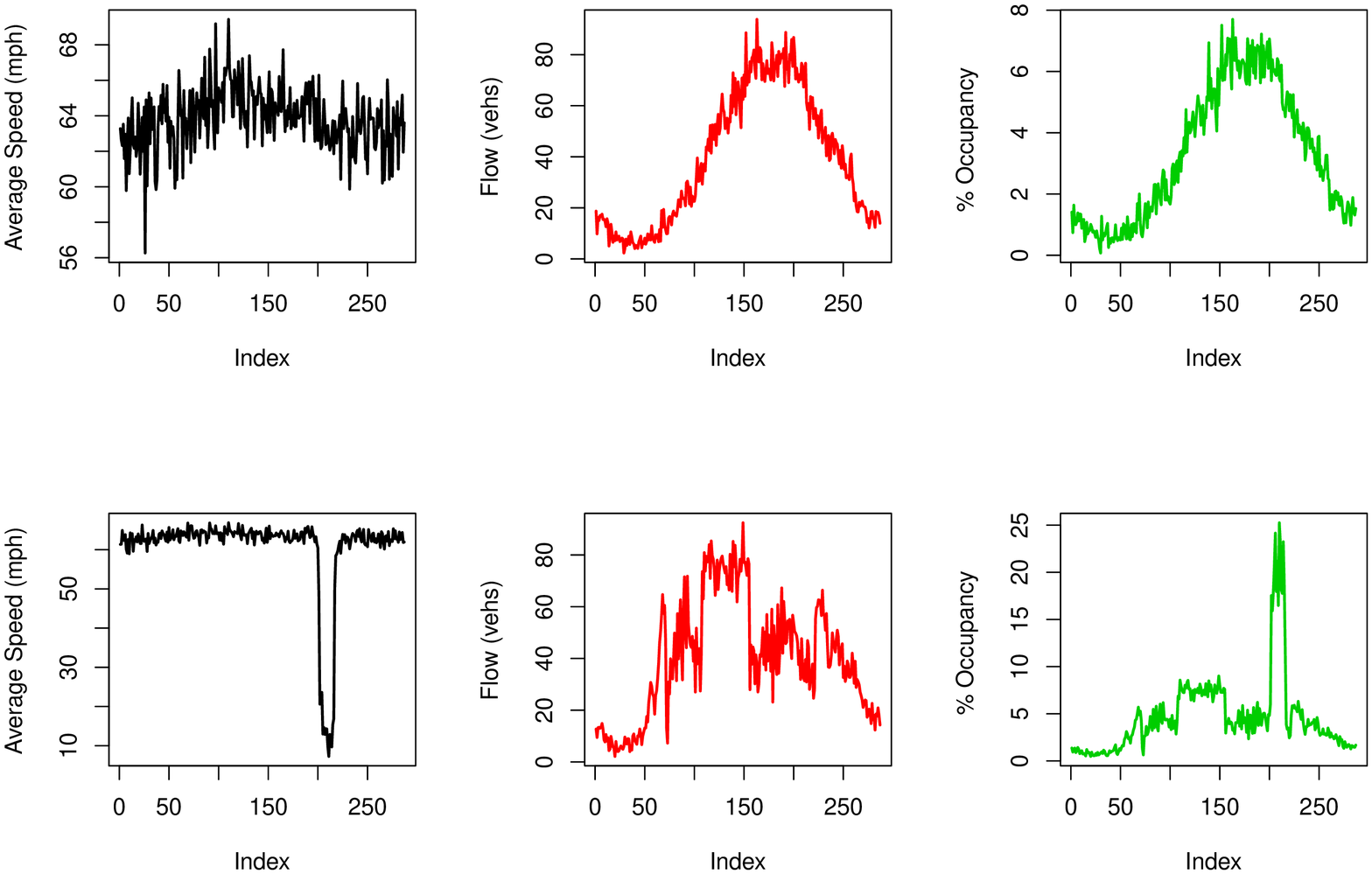}
\caption{Example 5-min Portland dataset normal vs accident days}
\label{fig_port}
\end{figure}

\begin{figure}[!tbp]
\centering
  \includegraphics[width=0.99\linewidth]{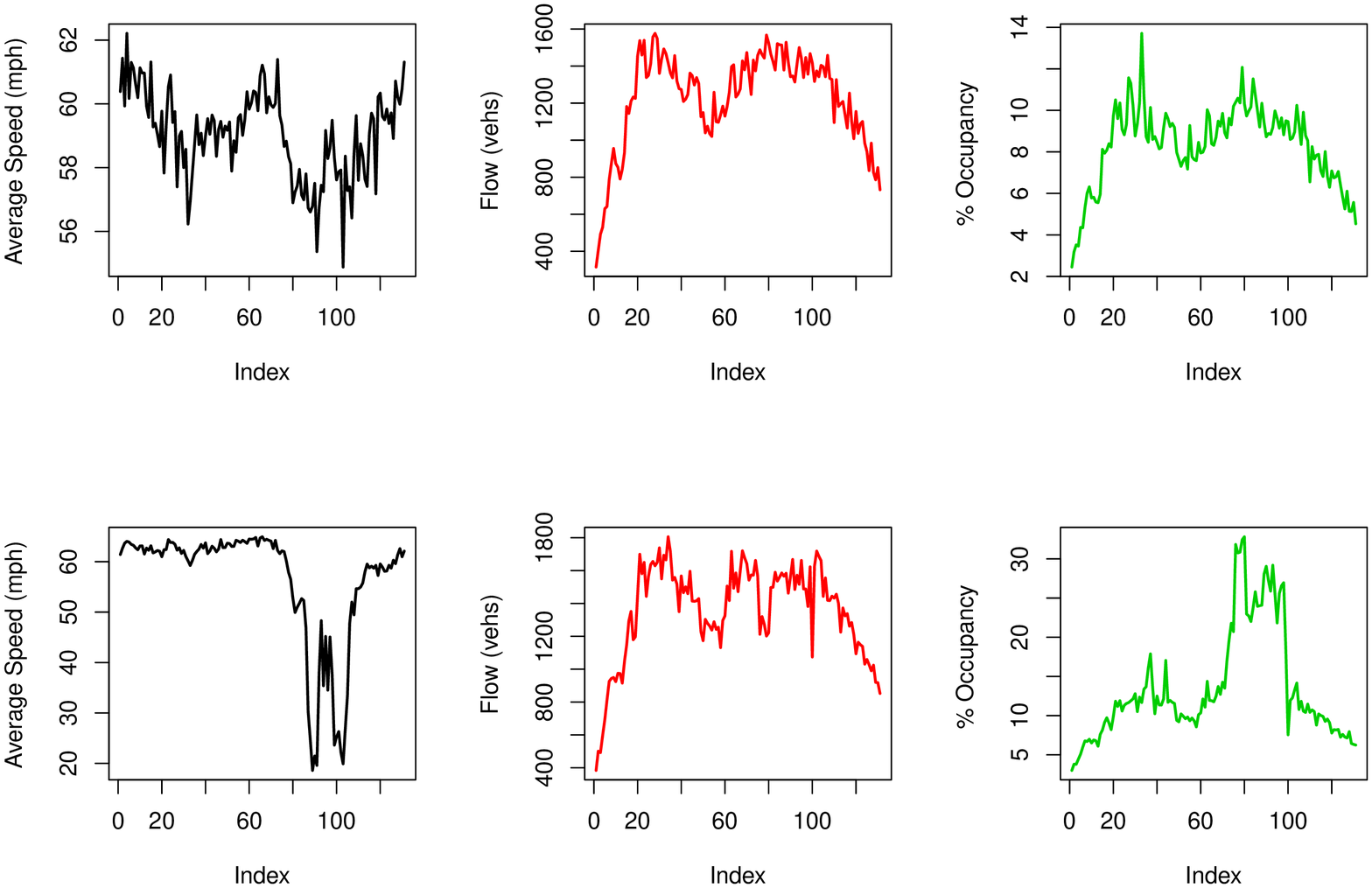}
\caption{Example 5-min CA dataset normal vs accident days}
\label{fig_cal}
\end{figure}

In Figs.~\ref{fig_port} and \ref{fig_cal}, it can be seen that normal days showing very similar values in both flow and density meaning vehicles are being discharged from a certain point without a bottleneck increased flow reflected as relatively increased occupancy (density) measured from the detector. However when we observe a sudden drop in speed, we can see that block shows reduced flow and increased occupancy of detector as vehicles would spend more time on the detector (e.g., stop and go like traffic). This behavior is similarly observed in CA dataset example in Fig.~\ref{fig_cal}. Thus, if one interested in predicting during dynamic regimes, if occupancy data is also available, it is better to use lagged speed and occupancy observations to predict average traffic speed.   

\section{Conclusions}
\label{sctconc}
This paper introduce univariate and multivariate Gaussian processes for one-step traffic speed forecasts at different aggregation levels (1-to-60 minutes) and compare them to Grey system models, and linear and nonlinear time series models. As the experiments demonstrate, the GP models provide consistently better performance and more accurate forecasts on all tested sample datasets. GP models are able to outperform the best performing GM model, i.e., Grey Verhulst model with Fourier error, on dynamic time series (e.g., time series involving multiple change points). As the experiments demonstrated, if another macroscopic traffic parameter is observed, using it along with lagged speed observations can also increase the accuracy. In fact, GP models also achieve better accuracy for the mid- and long-term predictions after aggregating the data. Key advantages of GP models are given below:

\begin{enumerate}
\item They only assume a covariance structure where linear kernel found to be effective in this study compared to quadratic and matern kernels;
\item GP models are found to be more accurate in short-term traffic parameter prediction against compared Grey systems, linear, and nonlinear models;
\item Among the GP models, multivariate model demonstrates the best accuracy and resistance to abrupt speed changes;
\item GP models can be better transferred to different datasets with different spatial and temporal behaviors; 
\item The computational times of the Gaussian Process models would allow the forecast algorithms to be placed on portable electronic devices. Although training times are relatively higher between 1-4 seconds, real-time predictions are possible as testing takes less 0.2 second.
\end{enumerate} 

Further work may include mid-term and long-term multi-step forecasts, their modifications to the basic equations, and a study on applicability of Gaussian Processes.
\section*{Acknowledgments}
This research is supported by U.S. Department of Homeland Security Summer Research Team Program for Minority Serving Institutions and follow-on grants managed by ORAU and was conducted at Critical Infrastructure Resilience Institute, Information Trust Institute, University of Illinois, Urbana-Champaign. The research also partially supported by U.S. Department of Transportation Regional Tier 1 University Transportation Center for Connected Multimodal Mobility ($C^2M^2$). It is also partially supported by NSF Grants Nos. 1719501, 1436222, 1954532, and 1400991.

\ifCLASSOPTIONcaptionsoff
  \newpage
\fi



%


\bibliographystyle{IEEEtran}
\bibliography{Transport_Bibliographytrb2}

%






\end{document}